\documentclass[journal]{IEEEtran}
\IEEEoverridecommandlockouts
\usepackage{graphicx}
\usepackage{array}
\usepackage{tabularx}
\usepackage{booktabs}
\usepackage{amsmath}
\usepackage{amssymb}
\usepackage{makecell}
\usepackage{float}
\usepackage{multirow}
\usepackage{stfloats}
\usepackage{algorithm} 
\usepackage{algorithmic} 
\usepackage{amsmath}
\usepackage{booktabs}
\usepackage{xcolor}    
\usepackage{etoolbox}  

\UseRawInputEncoding
\makeatletter
\newcommand{\removelatexerror}{\let\@latex@error\@gobble}
\makeatother
\usepackage{color}    

\ifCLASSINFOpdf
\else
\fi

\begin{document}
%
\title{Three-Stage Composite Outlier Identification of Wind Power Data: Integrating Physical Rules with Regression Learning and Mathematical Morphology}
%
%
%

\author{Limengqian~Zheng,
        Lipeng~Zhu,~\IEEEmembership{Member,~IEEE, }
        Weijia~Wen,
        Jiayong~Li,~\IEEEmembership{Member,~IEEE,}
        and Cong~Zhang,~\IEEEmembership{Member,~IEEE}
        
    \thanks{This work was supported in part by the National Natural Science Foundation of China under Grants 52207094, 52377184 and 52477090, in part by the Hunan Key Laboratory for Internet of Things in Electricity under Grant 2019TP1016, and in part by the Science and Technology Innovation Program of Hunan Province under Grant 2023RC3114 and Grant 2024RC3110. (Corresponding author: \textit{Lipeng Zhu}.)} 
    \thanks{Limengqian Zheng, Lipeng Zhu, Jiayong Li, and Cong Zhang are with the College of Electrical and Information Engineering, Hunan University, Changsha 410082, China (e-mail: a1479010379@163.com; zhulpwhu@126.com; j-y.li@connect.polyu.hk; zcong@hnu.edu.cn).}
    \thanks{Weijia Wen is with the State Grid Hunan Information \& Telecommunication Company, Changsha 410004, China (e-mail: cook\_lex@foxmail.com).}}

\maketitle

\begin{abstract}
Existing studies on identifying outliers in wind speed-power datasets are often challenged by the complicated and irregular distributions of outliers, especially those being densely stacked yet staying close to normal data. This could degrade their identification reliability and robustness in practice. To address this defect, this paper develops a three-stage composite outlier identification method by systematically integrating three complementary techniques, i.e., physical rule-based preprocessing, regression learning-enabled detection, and mathematical morphology-based refinement. Firstly, the raw wind speed-power data are preprocessed via a set of simple yet efficient physical rules to filter out some outliers obviously going against the physical operating laws of practical wind turbines. Secondly, a robust wind speed-power regression learning model is built upon the random sample consensus algorithm. This model is able to reliably detect most outliers with the help of an adaptive threshold automatically set by the interquartile range method. Thirdly, by representing the wind speed-power data distribution with a two-dimensional image, mathematical morphology operations are applied to perform refined outlier identification from a data distribution perspective. This technique can identify outliers that are not effectively detected in the first two stages, including those densely stacked ones near normal data points. By integrating the above three techniques, the whole method is capable of identifying various types of outliers in a reliable and adaptive manner. Numerical test results with wind power datasets acquired from distinct wind turbines in practice and from simulation environments extensively demonstrate the superiority of the proposed method as well as its potential in enhancing wind power prediction.
\end{abstract}


\IEEEpeerreviewmaketitle

\section{Introduction}
\subsection{Research Background}
\IEEEPARstart{A}{s} one of the most mature renewable energy sources, wind power has been rapidly developed over the past two decades due to its ubiquity, cleanliness, and renewability \cite{1}. However, the rapid increase of wind penetration levels also poses significant challenges to the safe and stable operation of power systems because of its intermittency, randomness, and volatility \cite{2}. To mitigate the adverse effects of wind power on power systems, it is essential to cope with the uncertainties of wind power by improving its predictability through the analysis of actual operational data. The operational data are predominantly acquired from the supervisory control and data acquisition (SCADA) systems, which continuously collect, monitor, and store the operating information of practical wind turbines \cite{99}, \cite{88}. Wind speed and power data, as the core information collected by SCADA systems, are extensively utilized for many advanced applications, e.g., wind power curve modeling \cite{3}, operational status monitoring \cite{4}, wind power prediction \cite{5}, etc. 
In practical SCADA systems with complex and imperfect sensing and data transferring conditions, wind speed and power data are often contaminated by a large number of outliers with erroneous measurements, which would significantly degrade the performance of SCADA data-enabled applications. Additionally, in the presence of some extreme or abnormal operating conditions, e.g., wind power curtailment,  the SCADA systems may acquire abnormal measurements with correct values yet significantly deviating from the standard wind speed-power data distribution. If not properly identified, these data may mislead SCADA data-driven applications to infer abnormal mapping relationships between the SCADA data and wind power outputs, thereby undermining the reliability of the corresponding applications like wind power prediction \cite{6}, \cite{qfxd}. Therefore, to ensure the reliability of SCADA data-driven applications, especially for wind power prediction, it is imperative to accurately identify outliers in the SCADA data. It should be noted that, though out of the scope of this paper,  it is crucial to further fix the identified outliers by data correction for data quality enhancement.
Considering that other meteorological factors like temperature and humidity have relatively low influences on SCADA data-enabled applications, this paper concentrates on identifying outliers in the specific SCADA data of wind speed-power pairs.

\subsection{Literature Review}
Up to now, numerous research efforts have been devoted to outlier identification of wind speed-power data, which can be broadly categorized into statistical analysis-based methods and image processing-based methods, as detailed in the following.

Statistical analysis-based methods primarily leverage the differences in statistical characteristics between normal and abnormal data for outlier identification \cite{iqrmethod}-[19]. Among these methods, the interquartile range (IQR) method \cite{iqrmethod} and the three-sigma rule \cite{3sigma} are the simplest and most widely used approaches to identify anomalous data. However, the three-sigma rule assumes that data follow a normal distribution, limiting its applicability in practice where field wind power data are irregularly distributed. Furthermore, given that abnormal data exhibit distinct statistical characteristics compared to normal data, outlier identification methods based on more complicated data distribution characteristics \cite{yfd}, \cite{knn} have recently gained significant attention. Key statistical characteristics include density, variance, variance change rate, etc. Some representative studies are reviewed below.

In \cite{ouyang}, Ouyang \textit{et al}. take a combination of statistical analysis and exponential smoothing to dynamically calculate the data's mean value. Data points with mean values exceeding a preset threshold are identified as anomalies. However, this method may fail to adapt to scenarios with an intensive concentration of stacked outliers (concentrated outliers with high density w.r.t. data distribution) and is prone to misdetection.
In \cite{bdfz}, Shen \textit{et al}. locate a specific type of point, referred to as a variation point, by analyzing the rate of change in power variance and applying the least squares method within each wind speed interval. Data beyond the variation point are then classified as anomalous. However, it is difficult for this method to address anomalous data in the upper-left region of the wind speed-power data space. Zheng \textit{et al}. \cite{8} introduce a density-based local outlier factor (LOF) algorithm to identify outliers by calculating the outlier factor for each data point. However, this method is highly sensitive to data distribution, which may experience performance degradation when there are a large number of outliers adjacent to normal data. Zhao \textit{et al}. \cite{9} integrate the density-based spatial clustering of applications with noise algorithm (DBSCAN) with the IQR method for outlier identification. They use the IQR method to detect dispersive outliers (scattered outliers with low density w.r.t. data distribution) and leverage DBSCAN to identify stacked outliers. However, the complexity of the parameter setting involved in this method poses some challenges, as improper setting may undermine the identification reliability. 
In \cite{otu}, Zhao \textit{et al.} stack autoencoders to learn high-dimensional features and calculate the reconstruction error for each data point. A threshold is determined by maximizing the variance between classes (i.e., normal data and stacked outliers), and data points with large errors are then identified as stacked outliers. Dispersive outliers are subsequently eliminated using DBSCAN.
However, this method has high computational costs and may incorrectly dismiss anomalies when the boundary between normal and anomalous data is complex or unclear. Wang \textit{et al}. \cite{wang2014copula} construct a joint probability distribution between wind speed and power through a Gaussian mixture copula model, and then identify and eliminate anomalies that significantly deviate from the joint distribution. However, a large number of stacked outliers may distort the model's estimation of the inherent wind speed-power data distribution, thereby degrading its performance on outlier identification.

On the whole, the aforementioned statistical methods \cite{ouyang}-\cite{wang2014copula} have exhibited their advantages in addressing outliers of wind power data, yet their realization relies on the assumption that the normal data dominate the dataset and their reliability is likely to be affected by the complicated procedures of empirical parameter and/or threshold settings.

In recent years, image processing-based methods have shown significant potential in the outlier identification of wind power data. These methods usually represent the wind speed-power distribution with a two-dimensional image, which is then utilized to detect outliers through advanced image processing techniques. 
Long \textit{et al}. \cite{hu} take mathematical morphological operations to extract the main components of the wind power image and calculate their Hu matrix related to the reference wind power curve. The method then optimizes the selection of structuring elements by minimizing the differences between the extracted components and the reference curve. Finally, it classifies data as normal or anomalous using an image edge detection technique.
Wang \textit{et al}. \cite{12} propose an image pixel-based outlier detection method. They utilize the distinct distributional characteristics of normal and anomalous data in the image to identify outliers through vertical and horizontal pixel cleaning. The largest continuous pixel segments are retained in each column and each row to remove the stacked outliers and the dispersive outliers, respectively. 
Su \textit{et al}. \cite{su} propose an image thresholding-based outlier identification method. This method separates the image into normal and abnormal data by applying global thresholding to the generated gray-level feature image. Liang \textit{et al}. \cite{13} initially set an outlier threshold by minimizing the dissimilarity and uncertainty-based energy (MDUE), then mark pixels with values below the threshold as outliers. These representative image processing-based methods \cite{hu}-\cite{13} can effectively handle stacked outliers, but they face the challenge of tuning the algorithm-specific parameters or thresholds, which incurs high computational costs.

\subsection{Motivation and Contribution}
To address these research gaps mentioned above, this paper develops a three-stage composite outlier identification method with a systematic integration of physical rules, regression learning (RL), and mathematical morphology (MM). First, a group of physical rules related to wind turbine operations is applied to eliminate anomalous data that conspicuously violate the operational principles. Then, a robust wind speed-power regression model is developed using the random sampling consensus algorithm (RANSAC) to detect most of the outliers with thresholds automatically set by the IQR method. Finally, from the data distribution perspective, mathematical morphology operation (MMO)-enabled image processing is applied to perform refined outlier identification. The primary contributions and advantages of this study are as follows.

\begin{enumerate}[]
\item This work develops a systematic data-driven scheme for outlier identification that integrates three heterogeneous methods to detect anomalous wind power data from complementary perspectives. With this integration, the proposed scheme is capable of fulfilling the challenging task of accurately identifying different types of outliers across various anomaly distributions; it also solves the troublesome issue of reliably separating high proportions of stacked outliers from neighboring normal data.

\item In contrast to conventional solutions with fixed outlier thresholds, the proposed scheme performs adaptive outlier detection with the help of the IQR method for automatic threshold setting. With no reliance on domain expertise or costly computational efforts for threshold determination, the proposed scheme can be implemented in a highly cost-effective manner, being more applicable than the former in practical contexts.

\end{enumerate}

The remaining sections of the paper are structured as follows. Section II presents the characteristics of outliers in wind speed-power data. Section III provides a comprehensive description of the proposed outlier identification methodology. In Section IV, numerical tests are conducted with several wind power datasets acquired from different wind turbines in practice and from simulation environments to extensively verify the performance of the proposed methodology. Finally, Section V concludes the paper.

\section{Characteristics of Outliers}
\begin{figure}[!t]
    \centering
    \includegraphics[width=1\linewidth]{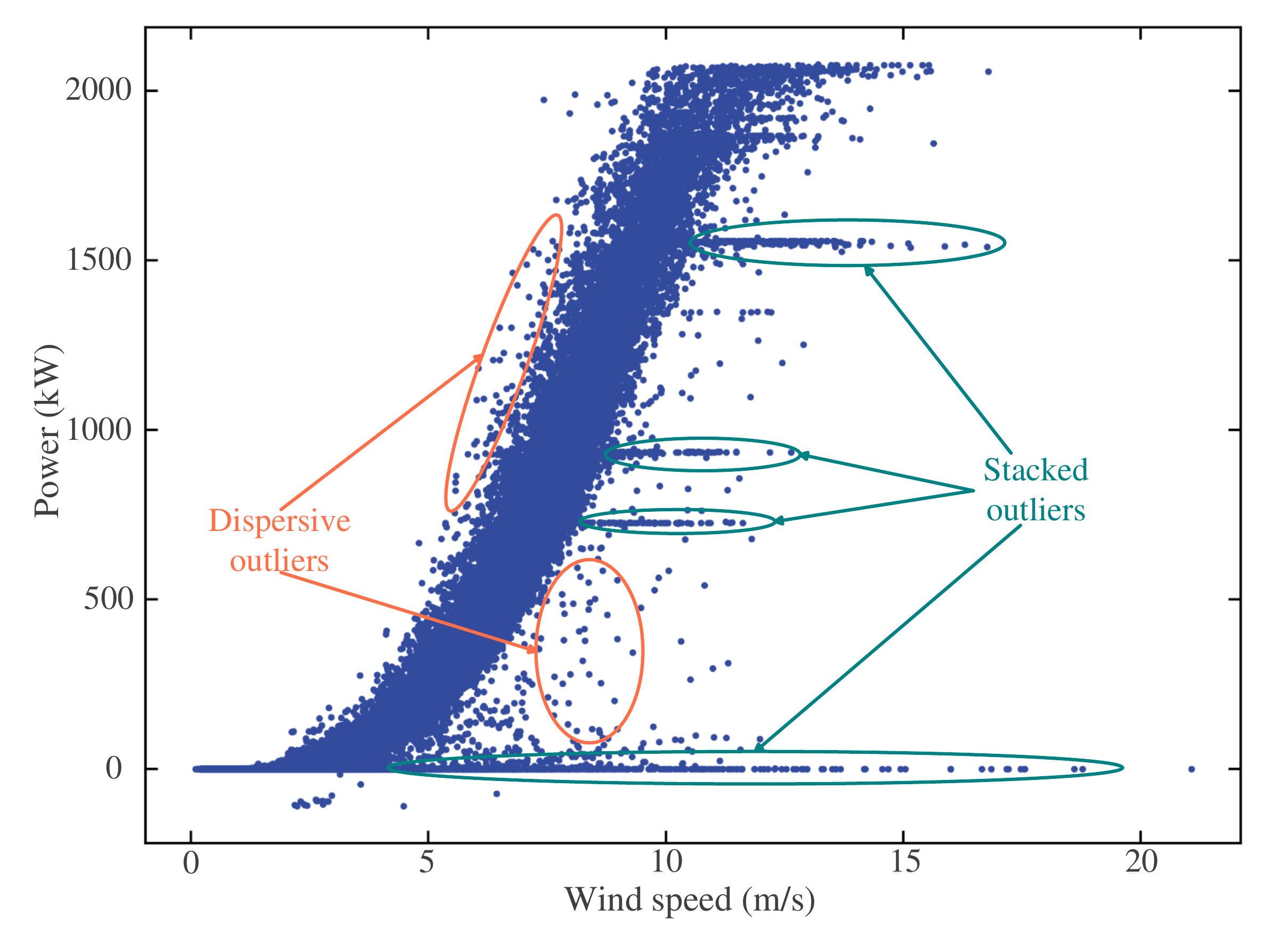}
    \caption{Distribution characteristics of outliers in wind power data.}
    \label{fig:enter-label1}
\end{figure}

With the increasing integration of wind power into modern power systems, the wind power curve, which visualizes wind power output under varying wind speed conditions, is essential for wind farm performance evaluation and power system operational optimization \cite{14}. However, the defective operating conditions of the SCADA systems in practical wind farms, e.g., sensor malfunctions, wind turbine failures, and data transmission errors, often introduce numerous outliers into the SCADA data, resulting in significant discrepancies between the actual and ideal wind power curves \cite{15}. Based on their distribution characteristics, outliers can be typically classified into dispersive outliers and stacked outliers, as shown in Fig. \ref{fig:enter-label1}. A brief overview of the characteristics of the two classes of outliers is provided in the following.

\subsection{Dispersive Outliers}
Dispersive outliers generally represent a number of scattered anomalies that are irregularly distributed in the wind speed-power data space, perhaps occurring at different time instants without fixed pattern or periodicity. These anomalies are generally caused by occasional sensor malfunctions, extreme weather, or uncontrolled random factors. Consequently, dispersive outliers do not present significant temporal or spatial correlations, and they are thus characterized by irregularity and dispersity.

\subsection{Stacked Outliers}
Stacked outliers are anomalies densely clustered in the data space. Such anomalies usually occur over a relatively long period, manifesting as patterned deviations of data points from normal data over time. The presence of stacked outliers may indicate a system-level operational problem during this time period, such as unplanned maintenance, wind turbine failures, measurement terminal failures, or wind power curtailment. Due to the prolonged duration and extensive influence, which may obscure their distinction from normal data, stacked outliers are more challenging to detect than dispersive outliers, requiring more sophisticated analysis and processing methods.

\section{Proposed Methodology}
\begin{figure}[!t]
    \centering
    \includegraphics[width=1\linewidth]{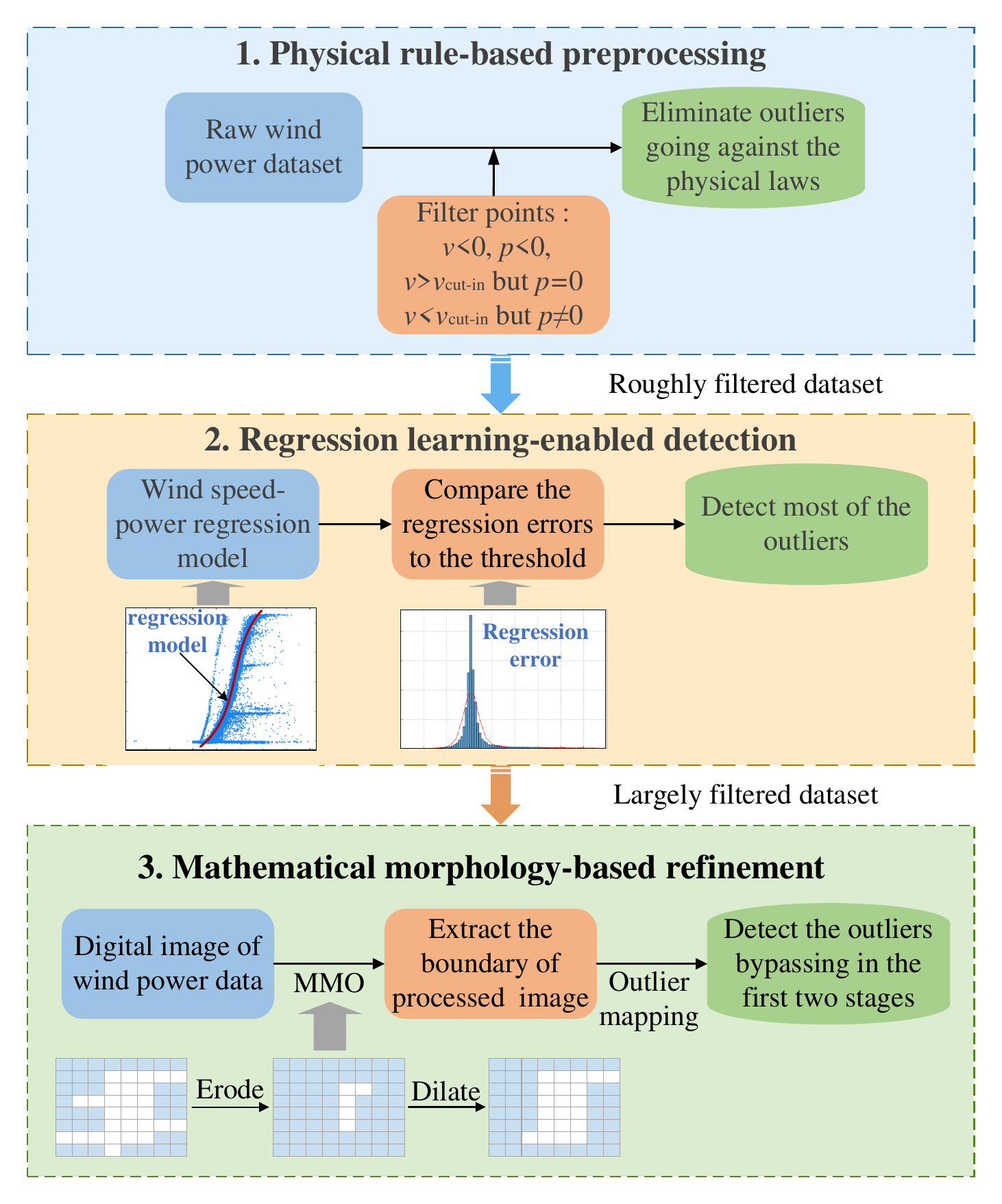}
    \caption{Overall framework of the proposed methodology.}
    \label{fig:enter-label0}
\end{figure}

In this section, the proposed outlier identification method is presented in detail. This method comprises three stages: physical rule-based preprocessing, RL-enabled detection, and MM-based refinement, as shown in Fig. \ref{fig:enter-label0}. The technical details are provided below.

\subsection{Physical Rule-Based Preprocessing}

In the initial stage, data points that satisfy one of the following rules are identified as outliers:

\begin{equation}v<0\end{equation}
\begin{equation}p<0\end{equation}
\begin{equation}v>v_\text{cut-in},p=0\end{equation}
\begin{equation}v<v_{\mathrm{cut-in}},p\neq0\end{equation}
where \textit{v} and \textit{p} represent the wind speed and power, respectively, and $v_\text{cut-in}$ denotes the cut-in wind speed. In practical wind turbine operations, a negative wind speed or a negative power output is counterintuitive, which violates fundamental physical principles. In addition, when the wind speed exceeds the cut-in speed, the wind turbine is expected to produce a certain amount of power based on the wind speed rather than zero output. Conversely, when wind speed falls below the cut-in speed, the generator lacks sufficient kinetic energy to produce electricity, and the power output should consequently be zero. Thus, operational data with abnormal characteristics represented by (1)-(4) can be directly marked as outliers.

In addition to those outliers going against fundamental physical rules, wind power datasets frequently involve missing and duplicate data. These data are considered a special form of outliers. To detect missing data, each entry in the dataset is examined for the presence of missing values, typically represented as NaN (not a number), null entries, or other designated numbers (e.g., -9999). For duplicate data, the key lies in determining whether the records are precisely identical across all relevant wind power features.

After filtering out the above obvious outliers, a large number of dispersive and stacked outliers still remain to be identified, and they will be handled in the following two stages.

 \subsection{ Regression Learning-Enabled Detection}
 
\begin{figure}
    \centering
    \includegraphics[width=1\linewidth]{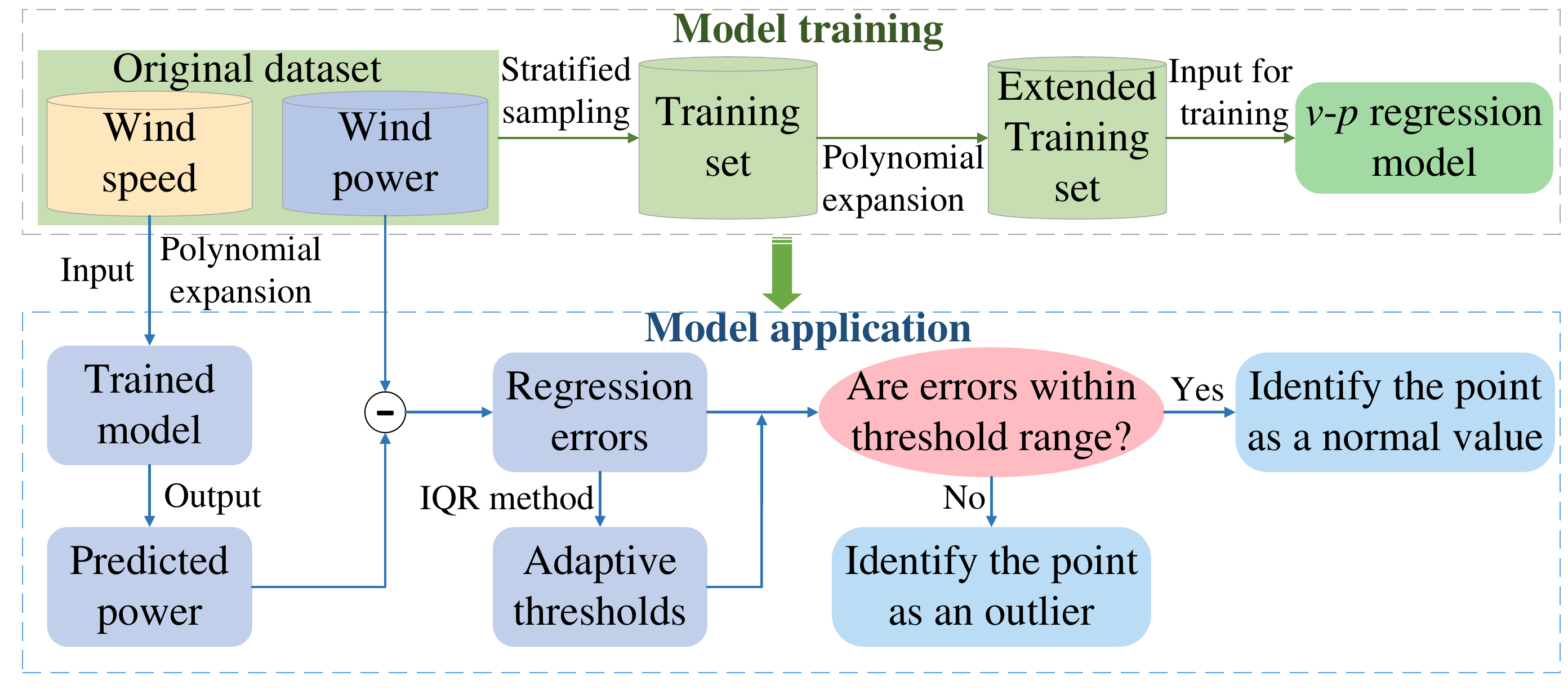}
    \caption{Overall process of RL-enabled outlier detection.}
    \label{RL-based}
\end{figure}

Considering the advantage of the RANSAC algorithm in enhancing the robustness of RL in the presence of anomalous data \cite{16}, a wind speed-power RL model is built upon this algorithm to extensively detect outliers.
The RANSAC algorithm is essentially an iterative technique used to estimate model parameters from datasets containing outliers. It categorizes data points into inliers and outliers: inliers are normal data points that can be described by the model and fall within the expected range, while outliers are anomalous data points that deviate significantly from the expected range \cite{17}. On the basis of the RANSAC algorithm, the overall process of RL-enabled outlier detection is shown in Fig. \ref{RL-based}.

In particular, the original wind power dataset is denoted as $\textit{\textbf{U}}=\{(v_1,p_1),\cdots,(v_i,p_i),\cdots,(v_n,p_n)\}$, where \textit{n} is the total number of data points in the dataset, \(v_{i}\) and \(p_{i}\) represent the wind speed and power of the \textit{i}th data point, respectively. 
First, the original dataset \textit{\textbf{U}} is stratified according to the output values, i.e., the wind power level, and a certain proportion of data points are randomly chosen from each stratum to form a dataset for RL, denoted by \(\textit{\textbf{U}}_{t}\). Then, a regression model is trained using wind speed data from \(\textit{\textbf{U}}_{t}\) as the raw input, denoted as \(\textit{\textbf{V}}_t=[v_1^{\prime},\cdots,v_i^{\prime},\cdots,v_m^{\prime}]\), and wind power data \(\textit{\textbf{P}}_t=[p_1',\cdots,p_i',\cdots,p_m']\) from \(\textit{\textbf{U}}_{t}\) as the output. To enhance the model's capability in capturing the nonlinear relationships between wind speed and power, the input is expanded using the polynomial expansion technique. In this study, cubic polynomial feature expansion is carried out. By applying a cubic polynomial feature expansion to each wind speed $v_{i}^{\prime}$, the wind speed vector is expanded as $\mathbf{\textbf{\textit{x}}}_i=[1,v_i',(v_i')^2,(v_i')^3]$. Consequently, the raw wind speed data are transformed into an expanded matrix \textbf{\textit{X}}:

\begin{equation}\mathbf{\textbf{\textit{X}}}=\begin{bmatrix}1&v_1'&(v_1')^2&(v_1')^3\\1&v_2'&(v_2')^2&(v_2')^3\\\vdots&\vdots&\vdots\\1&v_i'&(v_i')^2&(v_i')^3\\\vdots&\vdots&\vdots\\1&v_m'&(v_m')^2&(v_m')^3\end{bmatrix}\end{equation}

Taking \textbf{\textit{X}} and $\textit{\textbf{P}}_t$ as the input-output pairs, a linear regression model based upon the RANSAC algorithm is built for robust regression modeling. The specific model building process is detailed as follows.

\textit{Step 1}. The smallest data subset needed to fit the regression model is randomly chosen, and it is denoted by $(v_i^{\prime},p_i^{\prime})$, $(v_j^{\prime},p_j^{\prime})$, $(v_l^{\prime},p_l^{\prime})$, and $(v_k^{\prime},p_k^{\prime})$. Then, the regression parameter $\boldsymbol{\theta} = \begin{bmatrix} \theta_0, \theta_1, \theta_2, \theta_3 \end{bmatrix}^\mathrm{T}$ is subsequently calculated as follows:

\begin{equation}\boldsymbol{\theta}=\begin{bmatrix}\theta_0\\ \theta_1\\ \theta_2\\ \theta_3\end{bmatrix}=\begin{bmatrix}1 & v_{i}^{\prime} & (v_{i}^{\prime})^2 & (v_{i}^{\prime})^3\\ 1 & v_{j}^{\prime} & (v_{j}^{\prime})^2 & (v_{j}^{\prime})^3\\ 1 & v_{l}^{\prime} & (v_{l}^{\prime})^2 & (v_{l}^{\prime})^3\\ 1 & v_{k}^{\prime} & (v_{k}^{\prime})^2 & (v_{k}^{\prime})^3\end{bmatrix}^{-1}\begin{bmatrix}p_{i}^{\prime}\\ p_{j}^{\prime}\\ p_{l}^{\prime}\\ p_{k}^{\prime}\end{bmatrix}\end{equation}

\textit{Step 2}. Based on Step 1, the residual $r_{i}$ for each data point $(v_i^{\prime},p_i^{\prime})$ in the dataset is calculated:

\begin{equation}r_i=p_i^{\prime}-(\theta_0+\theta_1v_i'+\theta_2(v_i')^2+\theta_3(v_i')^3)\end{equation}
Next, the residual $\textit{\textbf{r}}=[r_1,r_2,\ldots,r_n]$ is compared with a given threshold \textit{h}. Data points with residuals less than \textit{h} are classified as inliers, while the remaining ones with residuals larger than \textit{h} are taken as outliers. \textit{h} is generally determined by (8):

\begin{equation}h=c\cdot M_{e}\left(\{|r_1-\tilde{r}|,\ldots,|r_{i}-\tilde{r}|,\ldots,|r_{n}-\tilde{r}|\}\right)\end{equation}
where \textit{c} is a constant factor, $\tilde{r}$ represents the median of all residuals, and $M_{e}(\textit{\textbf{x}})$ in this context refers to a function that estimates the mean value of all the elements in \textit{\textbf{x}}.

\textit{Step 3}. By repeating Steps 1 and 2 multiple times, the regression parameter yielding the highest count of inliers at each iteration is recorded until it reaches the maximum iteration number ${N_{\mathrm{max}}}$.

\textit{Step 4}. After completing all the iterations, the regression parameter obtained from the iteration with the largest number of inliers is selected to build the eventual regression model. The inlier set $\mathcal{I}$ is defined as follows:

\begin{equation}\mathcal{I}=\{(v_i^{\prime},p_i^{\prime})\mid r_i<h\}\end{equation}
Subsequently, the final model parameter $\boldsymbol{\theta}^{*}$ is estimated as:
\begin{equation}\boldsymbol{\theta}^*=\arg\max_\theta|\mathcal{I}(\boldsymbol{\theta})|\end{equation}

Note that, according to the authors' empirical tests, the RL-enabled outlier detection performance remain at satisfactory levels when the iteration number ${N_{\mathrm{max}}}$ and the constant factor \textit{c} for determining \textit{h} are set to typical values by default. In this regard, there is no need to perform time-consuming parameter tuning for them. Therefore, unless otherwise stated, ${N_{\mathrm{max}}}$ and \textit{c} are empirically set to 1000 and 1.43 in the sequel.

Next, the wind speed data in \textit{\textbf{U}} are expanded via cubic polynomial feature expansion [similar to the wind speed data in (5)], and the expanded wind speed data are input into the trained model to estimate wind power outputs. Then, each wind power error is calculated as $e_i=\hat{p}_i-p_i$, where \(e_{i}\) represents the error between the predicted power \(\hat{p}_{i}\) and the actual power \(p_{i}\) for the \textit{i}th data point.
 
Then, a pair of error thresholds \(\textbf{\textit{t}}=\{t_{low},t_{up}\}\) can be appropriately determined via the IQR method \cite{iqr}, which effectively reduces the distortion in threshold determination induced by outliers that disproportionately affect the error distribution. The specific calculations are as follows:

\begin{equation}I_{Q R}=Q_3-Q_1\end{equation}
\begin{equation}t_{low}=Q_1-\textit{k}I_{QR}\end{equation}
\begin{equation}t_{up}=Q_3+\textit{k}I_{QR}\end{equation}
where $t_{low}$ and $t_{up}$ are the lower and upper boundary thresholds, respectively. \(Q_1\) and \(Q_3\) represent the first and third quartiles of the errors, respectively, and \textit{k} is an adjustment factor. With \(Q_1\) and \(Q_3\) determined by the inherent numerical distribution of the dataset, the resulting thresholds $t_{low}$ and $t_{up}$ can adapt well to different datasets, thereby gaining desirable adaptability.

Finally, each point's error is compared with the thresholds $t_{low}$ and $t_{up}$ to identify anomalous and normal data points, which are further collected as an anomalous dataset \(\textit{\textbf{U}}_{a}\) and a normal dataset \(\textit{\textbf{U}}_{n}\):

\begin{equation}
\begin{aligned}
\textbf{\textit{U}}_{a} &= \{(v_{i}, p_{i}) \mid (v_{i}, p_{i}) \in \textit{\textbf{U}}, e_{i} \leq t_{low} \text{ or } e_{i} \geq t_{up}\} \\
\textit{\textbf{U}}_{n} &= \{(v_{i}, p_{i}) \mid (v_{i}, p_{i}) \in \textit{\textbf{U}}, t_{low} < e_{i} < t_{up}\}
\end{aligned}
\end{equation}

It is worth noting that, to comprehensively evaluate the outlier identification results in a reasonable manner, \textit{K}-fold cross validation \cite{K-fold} can be introduced in the RL process. Specifically, for each fold in the cross validation process, the data within each stratum are randomly divided into \textit{K} equal parts after stratifying \textit{U}. \textit{K}-1 subsets from each stratum are then randomly chosen to form \(\textit{\textbf{U}}_{t}\). Afterwards, the RL model is trained and applied to identify anomalous data from the whole dataset. Upon the completion of all folds, the results obtained from each fold undergo further refined outlier identification in the next stage. The eventual outlier identification performance derived from each fold is averaged to estimate the overall capability of the proposed method in anomaly detection. Essentially, the overall results validated in this way can be deemed as average results with \textit{K} times of repetitive experiments.

\subsection{Mathematical Morphology-Based Refinement}

To further improve the accuracy and robustness of outlier identification, the MM technique is introduced to finely identify the remaining outliers. 

MM is a robust digital image processing technique that leverages a structuring element to perform a range of morphological operations \cite{18}. By modifying the size and shape of the structuring element, precise control over image processing effects can be achieved. Typical structuring elements include rectangular, cross-shaped, and circular forms. Within the domain of mathematical morphology operation, dilation and erosion serve as the fundamental operations. When applied to digital images derived from wind power data, these operations facilitate the extraction of morphological features related to wind power curves \cite{19}, thus aiding in the identification of anomalous data. The algorithmic process of the MM-based refinement is depicted in Fig. \ref{fig:enter-label6}, with the detailed steps presented below:

\begin{figure}[!t]
    \centering
    \scalebox{0.9}{\includegraphics[width=1\linewidth]{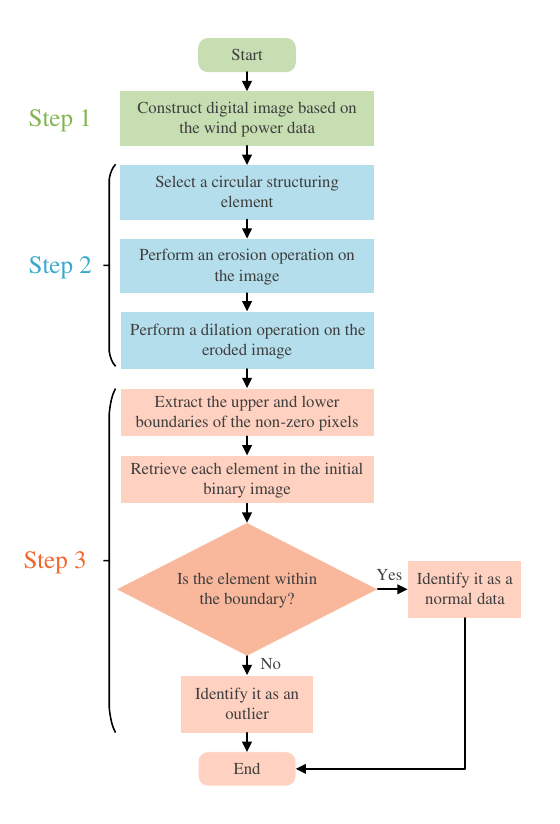}}
    \caption{Flowchart of the MM-based refinement.}
    \label{fig:enter-label6}
\end{figure}

\textit{Step 1}. With some outliers filtered out by the above two stages, the remaining data points are gathered as a subset $\textit{\textbf{U}}_{r}$ and represented by a digital image in the wind speed-power data space.

1) Each data point in $\textit{\textbf{U}}_{r}$ is normalized as follows: 

\begin{align}
& \bar{v}_i=\frac{v_i-v_{\min }}{v_{\max }-v_{\min }} \\
& \bar{p}_i=\frac{p_i-p_{\min }}{p_{\max }-p_{\min }}
\end{align}
where \(v_i\) and \(p_i\) are the wind speed and power of the \textit{i}th data point in $\textit{\textbf{U}}_{r}$. $v_{\min }$, $v_{\max }$, $p_{\min }$, and $p_{\max }$ denote the minimum and maximum values of the wind speed and power, respectively. 
 
2) The normalized data are then converted to a specific range of positive integers:

\begin{align}
&\tilde{\nu}_{i}=f(\overline{\nu}_{i} q)+1\\
&\tilde{p}_{i}=f(\overline{p}_{i} q)+1\\
&f(x)=\begin{cases}[x], x\geq0\\ [x]+1, x<0\end{cases}
\end{align}
where \textit{q} is a positive integer that determines the range of the resulting integer values, and \textit{f}(\textit{x}) is a floor function, expressed as [19], with $[x]$ denoting the largest integer less than or equal to $x$. To ensure uniform data coverage while preserving a sufficient level of detail in the image, \textit{q} is set to a typical value of 100. This typical value facilitates an efficient distribution of data points, preserving the granularity of the data while minimizing the risk of excessive overlap or information loss.

3) A two-dimensional matrix $\textit{\textbf{M}}=[m_{i,j}]_{(q+1)\times(q+1)}$ is initialized, where all the entries are initialized as $\mathrm{\textit{m}_{\textit{i},\textit{j}}=0}$. Then, all data points in $\textit{\textbf{U}}_{r}$ are mapped to \textit{\textbf{M}} by indexing the matrix elements using the coordinates ${\left(\tilde{\nu}_{\mathrm{i}},\tilde{\mathrm{p}}_{\mathrm{i}}\right)}$, and setting the element $\mathrm{\textit{m}}_{\tilde{\nu}_i,\tilde{p}_i}$ to 1. Afterwards, each element of the matrix is mapped to its corresponding pixel value, with 1 represented by a black pixel and 0 represented by a white pixel. In this way, \textit{\textbf{M}} is converted into a digital image, denoted by \textit{A}.

\textit{Step 2}. MMOs are applied to \textit{A}. 

1) A circular structuring element, denoted as \textit{B}, with its diameter \textit{d}, is selected and employed to perform an erosion operation on \textit{A}, producing the eroded image $A_{E}$:

\begin{equation}A_E=A\ominus B=\{z\mid(B)_z\subseteq A\}\end{equation}
In this process, \textit{B} continuously moves over \textit{A}. At each pixel, it checks whether \textit{B} is entirely contained within the pixel set of \textit{A}. If this condition is satisfied, the pixel value at that position is set to 1. This operation effectively shrinks the image, removing noise and reducing the size of objects within the image.

2) \textit{B} is then utilized to execute a dilation operation on $A_{E}$, thereby resulting in the dilated image $A_{D}$:

\begin{align}
A_D=A_{E} \oplus B & = \left\{z \mid\left(B^{R}\right)_{z} \cap A_{E} \neq \varnothing\right\}
\end{align}
Specifically, the reflected structuring element \(B^{R}\) moves pixel by pixel across $A_{E}$, checking for overlap at each position. If an overlap is detected, the pixel value at that position is set to 1. This process helps to fill pores and connect broken structures.

\textit{Step 3}. With $A_{D}$ derived from Step 2, refined outlier detection is performed as follows.

1) Each column of $A_{D}$ is traversed to extract the coordinates of the upper and lower boundaries of the non-zero pixels. These boundary coordinates are then interpolated to generate a smooth envelope.

2) Then, each data point in \textit{A} is examined to assess whether it resides within the envelope. If a point falls outside the envelope, its index is recorded, and the corresponding data point in $\textit{\textbf{U}}_{r}$ is marked as an outlier.

\subsection{Outlier Identification Performance Evaluation}
To comprehensively examine the efficacy of the proposed methodology, its performance is tested from the following three different aspects.

\subsubsection{Wind Power Curve Fitting Error}
First, since the wind power curve fitting error is widely used in the research community to quantify outlier identification performance \cite{9}, \cite{12}, \cite{Luo11}, it is chosen as the straightforward metric to initially assess the potential of the proposed methodology in removing outliers far from the normal wind power curve.
In this study, wind power curves are obtained using the 'bin' method, which results in discrete curves \cite{20}. Cubic spline interpolation \cite{dxscz} is then applied to derive continuous curves. With continuous wind power curves, the root mean square error (RMSE) is used to quantify the error in wind power curve fitting:

\begin{equation}E_{\mathrm{rmse}}=\frac{1}{C}\sqrt{\frac{1}{N}\sum_{i=1}^{N}\left(\hat{p}_{i}-p_{i}\right)^{2}}\end{equation}
where \textit{C} is the rated power, \textit{N} is the number of power data points, $p_{i}$ and $\hat{p}_i$ represent the actual power value at a certain wind speed and the corresponding power value on the wind power curve at the same wind speed, respectively. A smaller $E_{\mathrm{rmse}}$ indicates a lower error in wind power curve fitting.

\subsubsection{Classification Potential with Indisputable Labels}

For labeled datasets, in addition to the wind power curve fitting error, classification-based indices are used to assess outlier identification performance. Specifically, since both normal and abnormal data points are labeled, accuracy ($A_{\mathrm{cc}}$), error rate ($E_{\mathrm{rr}}$), and F1 score ($F_{\mathrm{1}}$) are used as evaluation metrics to directly estimate the performance of the proposed outlier identification methodology:

\begin{equation}A_{\mathfrak{cc}}=\frac{T_P+T_N}{T_P+T_N+F_P+F_N}\times100\%\end{equation}

\begin{equation}E_{\mathrm{rr}}=\frac{F_{P}+F_{N}}{T_{P}+T_{N}+F_{P}+F_{N}}\times100\%\end{equation}

\begin{equation}F_1=\frac{2PR}{P+R}\end{equation}
where $T_{P}$ and $T_{N}$ represent the number of anomalies and normal data which are correctly identified, respectively, $F_{P}$ indicates the number of normal data which are incorrectly identified as anomalies, and $F_{N}$ refers to the number of anomalies which are incorrectly identified as normal data. \textit{P} and \textit{R} represent precision and recall, respectively, and are formulated as $P=\frac{T_{P}}{T_{P}+F_{P}}$ and $R=\frac{T_{P}}{T_{P}+F_{N}}$.

\subsubsection{Effect of Outlier Identification on Enhancing Data Quality}

With the ultimate target of outlier identification set to providing a high-quality wind power dataset for follow-up applications like wind power prediction, the datasets before and after outlier identification (with the cubic spline interpolation technique adopted to fix outliers) are utilized for wind power prediction. By testing whether the prediction errors [mean absolute error (MAE) and RMSE] can be effectively reduced after outlier detection and correction, the effect of the proposed methodology on data quality enhancement is assessed as:

\begin{equation}e_{\mathrm{mae}}=\frac{1}{C} \frac{1}{n}\sum_{i=1}^{n}\mid y_i-\hat{y}_i\mid \end{equation}
\begin{equation}e_{\mathrm{rmse}}=\frac{1}{C}\sqrt{\frac{1}{n}\sum_{i=1}^{n}\left(y_{i} -\hat{y}_{i}\right)^{2}}\end{equation}
where \textit{n} is the total number of data points, $y_{i}$ and $\hat{y}_i$ are the actual and predicted power values of the \textit{i}th data point, respectively. The smaller $e_{\mathrm{mae}}$ and $e_{\mathrm{rmse}}$, the better the wind power prediction and outlier identification performance.

\section{Case Study}
\subsection{Data Description}
\begin{figure}[!t]
    \centering
    \includegraphics[width=1.0\linewidth]{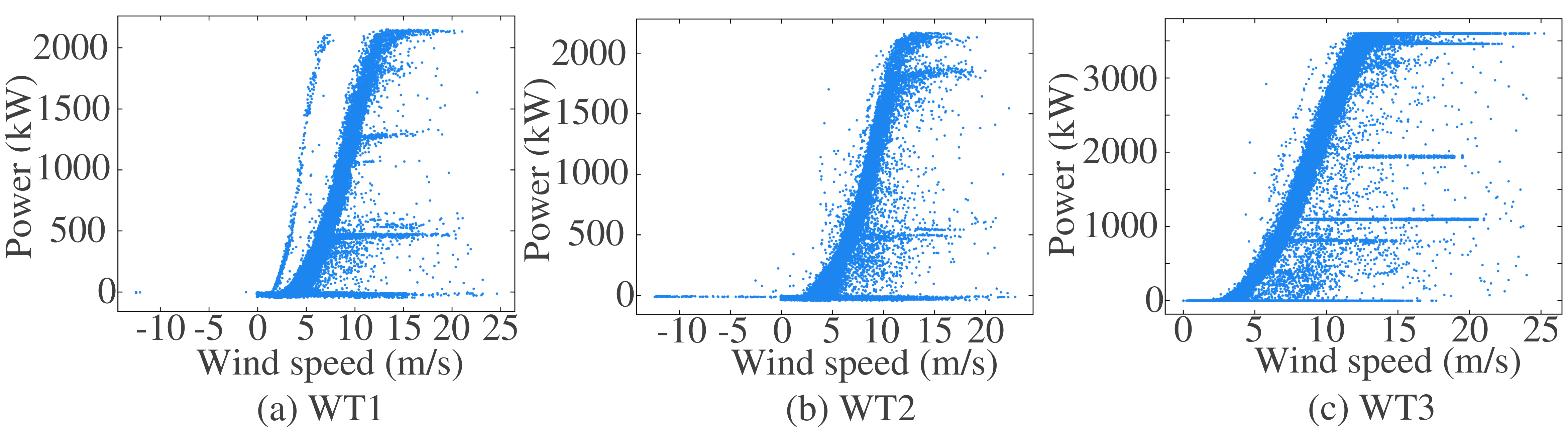}
    \caption{Wind power curve of each wind power dataset.}
    \label{fig:enter-label7}
\end{figure}

To verify the effectiveness of the proposed method, numerical tests are conducted using wind power SCADA datasets acquired from three actual wind turbines in three different wind farms, denoted as WT1, WT2 \cite{cnwind}, and WT3. Among them, wind speed-power data collected from WT1 and WT2 form unlabeled datasets (without outlier annotation), while wind speed-power data points acquired from WT3 constitute a labeled dataset (with outlier annotation). 
This labeled dataset is generated by manually introducing outliers to a completely normal dataset cleansed by domain experts. 
The operational parameters of each wind turbine are presented in Table \ref{tab2}, and the original wind power curves are illustrated in Fig. \ref{fig:enter-label7}.

As shown in Fig. \ref{fig:enter-label7}, the three datasets involve complicated and irregular outliers, with differing distributions of anomalies, providing fruitful data sources to validate the effectiveness of the proposed method.

\begin{table}[!t]
    \belowrulesep=0pt
    \aboverulesep=0pt
    \renewcommand{\arraystretch}{1.3}
    \caption{Operational Parameters of Each Wind Turbine}
    \label{tab2}
    \centering
    \begin{tabularx}{\linewidth}{>{\centering\arraybackslash}c  >{\centering\arraybackslash}X  >{\centering\arraybackslash}X  >{\centering\arraybackslash}X >{\centering\arraybackslash}X}
    \toprule
    \multirow{2}{*}{Dataset} & Rated power (kW)&Cut-in wind speed (m/s)& Cut-off wind speed (m/s) & Number of data points \\ 
    \midrule
    WT1    & 2000            & 3                      & 25                   & 43324     \\ 
    WT2    & 2000            & 3                      & 25                   & 38470   \\ 
    WT3    & 3600            & 3                      & 25                   & 57121   \\ 
    \bottomrule
    \end{tabularx}
\end{table}

\subsection{Comprehensive Outlier Identification Performance}

\subsubsection{Overall Performance of the Proposed Method}
First of all, how the proposed three-stage outlier identification method and its key techniques involved in the three stages (i.e., physical rule-based preprocessing, RL-enabled detection, and MM-based refinement) perform on the above three datasets is comprehensively examined. 
{Note that in the RL-enabled detection stage, 5-fold cross validation is introduced to verify the outlier identification results in a more reasonable manner.
Furthermore, the parameters \textit{k} and \textit{d} are initially set to 1.5 and 6, respectively, as suggested by the initial studies first proposing the IQR and MMO methods \cite{tukey}, \cite{mmos}. The corresponding thresholds related to RL-enabled detection for the three datasets are presented in Table \ref{tab3}.

\begin{table}[!t]
    \belowrulesep=0pt
    \aboverulesep=0pt
    \renewcommand{\arraystretch}{1.3}
    \centering
    \caption{Threshold Settings for RL-Enabled Outlier Detection}
    \label{tab3}
    \begin{tabularx}{\linewidth}{
      >{\centering\arraybackslash}c 
      >{\centering\arraybackslash}X 
      >{\centering\arraybackslash}X
    }
        \toprule
        Dataset & $t_{low}$ (kW) & $t_{up}$ (kW) \\
        \midrule
        WT1    & -145.1 & 192.3 \\
        WT2    & -126.7 & 164.1 \\
        WT3    & -210.8 & 312.1 \\ 
        \bottomrule
    \end{tabularx}
\end{table}

To illustrate the potential of individual techniques involved in the three stages in enhancing outlier identification, a series of tests are conducted by only adopting physical rule-based preprocessing, RL-enabled detection, and MM-based refinement. In addition, the proposed composite method adopting all three techniques is considered for tests. The results of these tests are comprehensively presented in Fig. \ref{8}, Fig. \ref{9}, and Fig. \ref{10}, corresponding to outlier identification in the three datasets, respectively.

\begin{figure}[!t]
    \centering
    \includegraphics[width=1\linewidth]{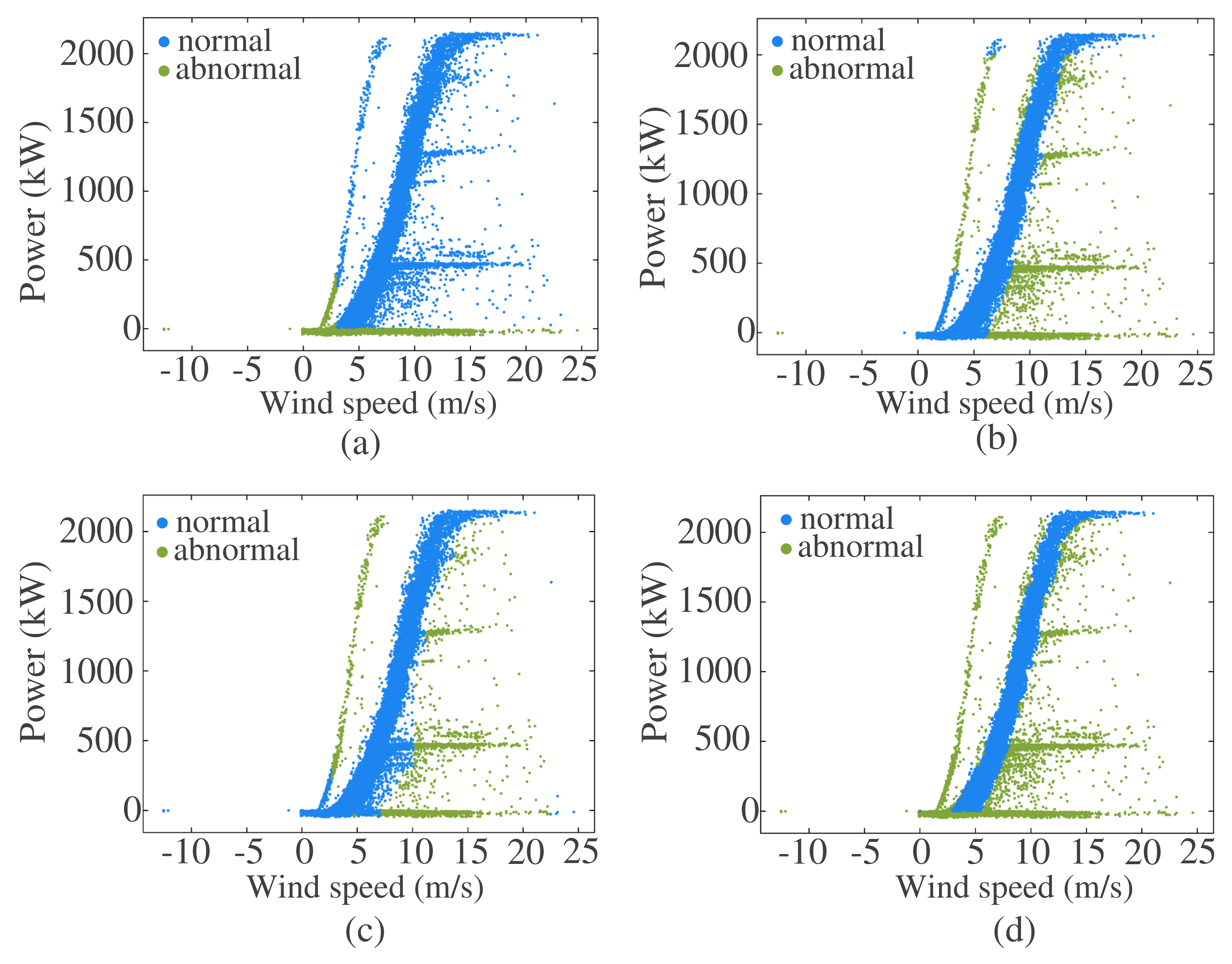}
    \caption{Outlier identification results for WT1. (a) Identification based on physical rules. (b) Identification based on RL. (c) Identification based on MM. (d) Identification based on the proposed method.}
    \label{8}
\end{figure}

\begin{figure}[!t]
    \centering
    \includegraphics[width=1\linewidth]{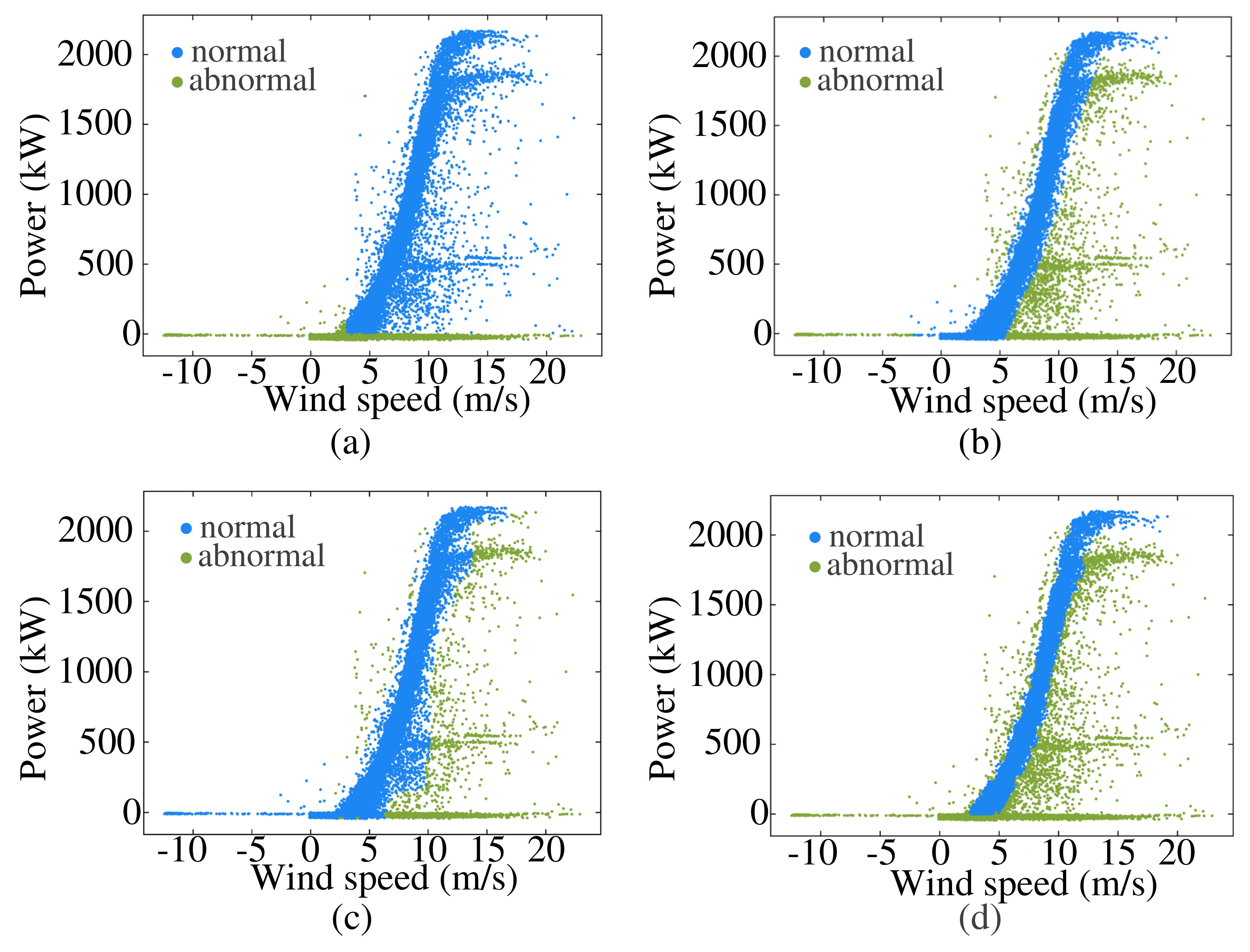}
      \caption{Outlier identification results for WT2. (a) Identification based on physical rules. (b) Identification based on RL. (c) Identification based on MM. (d) Identification based on the proposed method.}
    \vspace{-0.8em}
    \label{9}
\end{figure}

\begin{figure}[!t]
    \centering
    \includegraphics[width=1\linewidth]{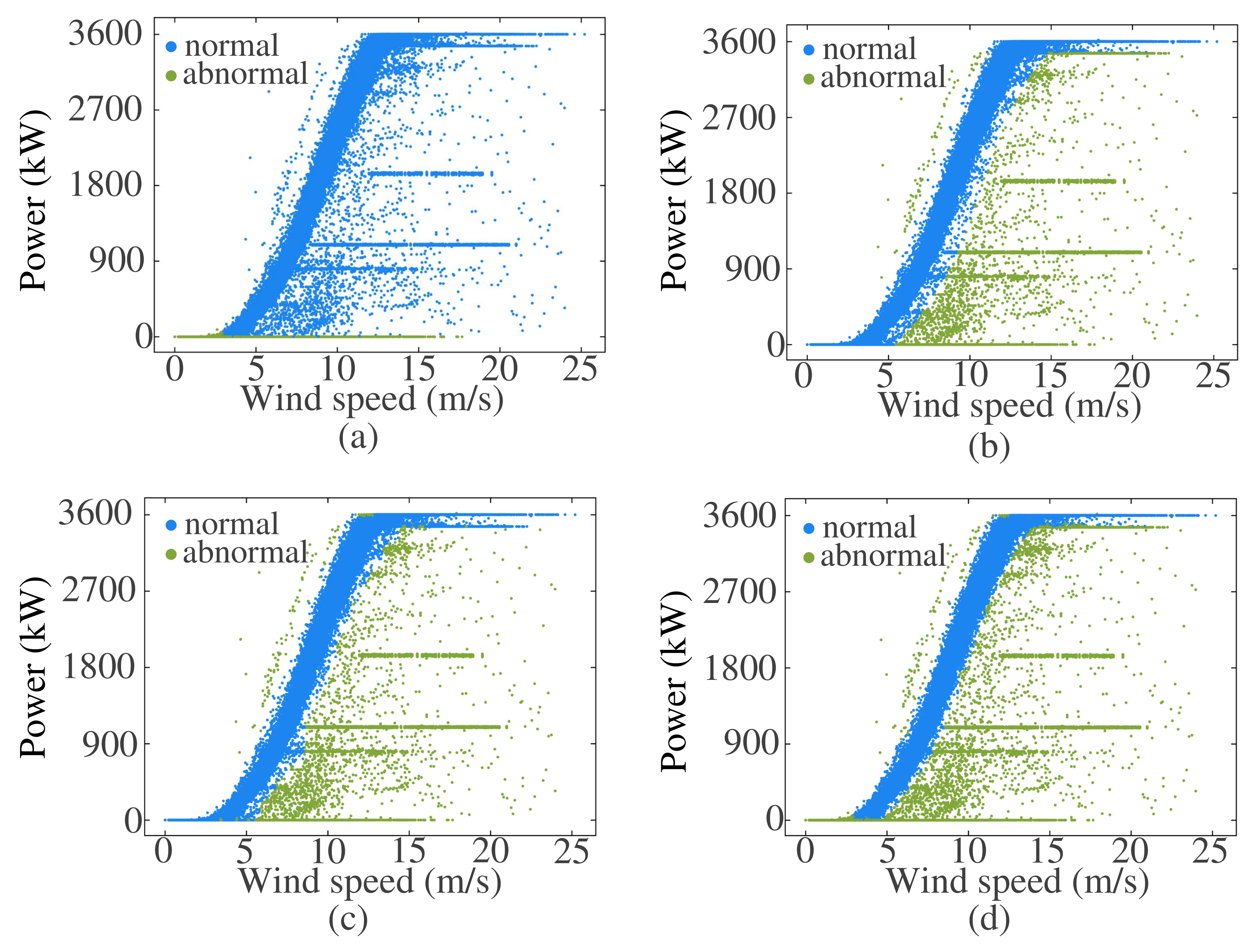}
    \caption{Outlier identification results for WT3. (a) Identification based on physical rules. (b) Identification based on RL. (c) Identification based on MM. (d) Identification based on the proposed method.}
    \label{10}
\end{figure}

As can be observed, the method based on physical rules effectively detects the stacked outliers in the bottom left corner of the wind speed-power space, yet a large number of anomalies in other positions remain to be detected. Then, it is found that considerable anomalies are successfully identified by the RL-based method and MM-based method, which implies the complementary nature of the three techniques. When the proposed composite method is applied, there is a significant enhancement in outlier identification compared to the only use of individual techniques. In Fig. \ref{8}(d), Fig. \ref{9}(d), and Fig. \ref{10}(d), the blue data points representing normal data highly resemble the ideal wind power curves, thus confirming the effectiveness and reliability of the proposed method regarding outlier identification. Additionally, across different datasets, the proposed method demonstrates superior outlier identification performance in spite of the patterns of outliers. This also indicates its strong generalization performance when applied to different wind farms in practice.

Furthermore, to provide a more intuitive comparison, Fig. \ref{r3} shows the scatter plots of the actual normal wind power generation data and the data cleansing for WT3, with potential outliers detected by the proposed method removed. As can be seen, the cleansed wind power curve derived from the proposed method closely resembles the original wind power curve. The identified normal data points in the wind speed-power plane do not simply resemble a single line but show a certain range of power variations at specific wind speed levels. This indicates that, after abnormal data removal, the remaining data points still reflect the natural relationship between wind speed and power, without distorting the inherent variability of the data. Therefore, the method proposed in this paper not only effectively removes abnormal data, but also retains the inherent characteristics of practical wind power datasets. Expectedly, this would be very helpful for subsequent data-driven applications (e.g., wind power prediction and fault diagnosis) built upon the exploration of practical characteristics within wind power datasets.

\begin{figure}[!t]
    \centering
    \includegraphics[width=1\linewidth]{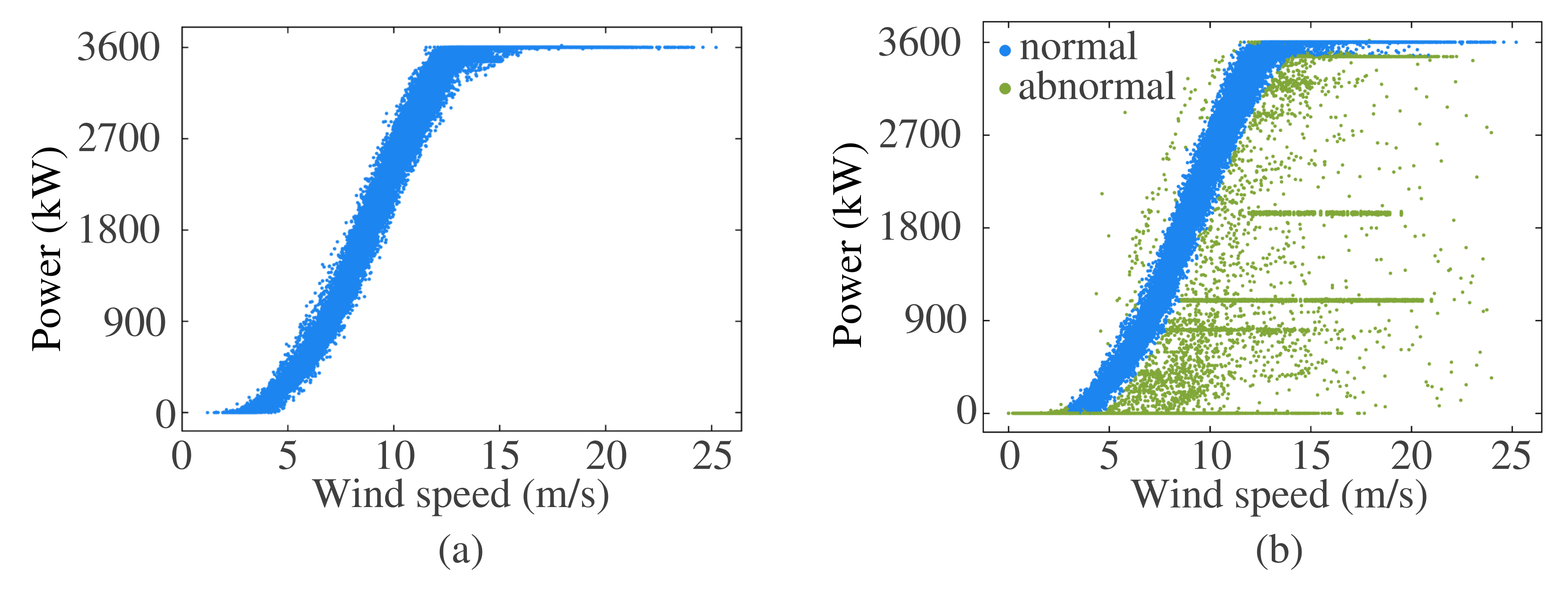}
    \caption{Comparison of wind power curves. (a) Actual normal wind power data points. (b) Cleansed wind power data points.}
    \label{r3}
\end{figure}

\subsubsection{Comparative Study}
To further validate the superiority of the proposed method, it is compared with some representative alternatives, including LOF \cite{8}, Quartile-DBSCAN \cite{9}, MDUE \cite{13}, the RL-based method, and the MM-based method for outlier identification. The performance of each method is summarized in Table \ref{tab888}. In addition, to visualize the performance of LOF, Quartile-DBSCAN, and MDUE, Fig. \ref{fig:enter-label9}, Fig. \ref{fig:enter-label10}, and Fig. \ref{fig:enter-label11} show their outlier identification results in the wind speed-power space.

As can be seen in Table \ref{tab888}, all methods manage to largely reduce the errors of wind power curve modeling. Overall, the proposed method defeats the other alternatives. It achieves superior accuracy and robustness for detecting anomalies distributed throughout the wind speed-power space across all datasets, with its $E_{\mathrm{rmse}}$ being no more than 0.037. 
In contrast, LOF exhibits the largest curve fitting errors, particularly in WT3, where the value of $E_{\mathrm{rmse}}$ is nearly 4.4 times higher than that of the proposed method. As observed in Fig. \ref{fig:enter-label9}, this method can only identify a subset of dispersive outliers, and it struggles to effectively detect clustered anomalies when they are densely concentrated in the middle of the wind speed-power space.
For the Quartile-DBSCAN-based method, significant variations are observed in wind power curve fitting errors across different anomaly distributions. As depicted in Fig. \ref{fig:enter-label10}, the stacked outliers within the central region of the wind power curves for WT1 and WT3 have not been effectively detected, resulting in $E_{\mathrm{rmse}}$ approximately 2 to 3 times higher than those of the proposed method. 
For the MDUE-based method, although it produces comparatively lower $E_{\mathrm{rmse}}$, Fig. \ref{fig:enter-label11} reveals that a significant amount of normal data near the rated power are misidentified as anomalous, leading to confusion between normal and anomalous data, particularly in the top region of the curve. The remaining RL-based and MM-based methods demonstrate commendable performance in wind power curve modeling, but they still do not perform as well as the proposed method, with more anomalies left undetected. The inferior performance of these comparative methods further implies the superiority of the proposed method.

Given the labeled dataset WT3, additional experiments are conducted to directly compare the effectiveness of different methods in outlier identification. Fig. \ref{fig:enter-label111} presents comparative results regarding $A_{\mathrm{cc}}$, $E_{\mathrm{rr}}$, and $F_{1}$ of outlier identification. Evidently, the proposed method outperforms others, achieving an accuracy of over 96\% and an F1 score of more than 0.94, along with the lowest $E_{\mathrm{rr}}$. Among the remaining methods, the RL-based and Quartile-DBSCAN-based methods also perform relatively well; however, their performance does not yet match that of the proposed method. Following these, the MM-based, LOF-based, and MDUE-based methods exhibit higher $E_{\mathrm{rr}}$. In particular, MDUE's $E_{\mathrm{rr}}$ reaches up to 20\%, resulting in the lowest $F_{1}$, which does not exceed 0.7. 

Based on the above analyses, it is evident that the proposed method achieves the best outlier detection performance, remaining robust to various practical outliers with different distributions.

\begin{table}[!t]
    \renewcommand{\arraystretch}{1.3} 
    \caption{Comparison of Different Methods for Outlier Detection ($E_{\mathrm{rmse}}$ in [p.u.])}
    \label{tab888}
    \centering
    \begin{tabularx}{\linewidth}{>{\centering\arraybackslash}c  
                                    >{\centering\arraybackslash}X  
                                    >{\centering\arraybackslash}X  
                                    >{\centering\arraybackslash}X  
                                    >{\centering\arraybackslash}X  
                                    >{\centering\arraybackslash}X  
                                    >{\centering\arraybackslash}X  
                                    >{\centering\arraybackslash}X}
    \toprule
    Dataset & Original dataset & Proposed method & LOF & Quartile-DBSCAN & MDUE & RL & MM \\
    \midrule
    WT1 & 0.141 & 0.035 & 0.105 & 0.103 & 0.044 & 0.047 & 0.055 \\
    WT2 & 0.133 & 0.036 & 0.058 & 0.037 & 0.038 & 0.046 & 0.045 \\
    WT3 & 0.175 & 0.031 & 0.141 & 0.072 & 0.032 & 0.033 & 0.033 \\
    \bottomrule
    \end{tabularx}
\end{table}

\begin{figure}[!t]
    \centering
    \includegraphics[width=1\linewidth]{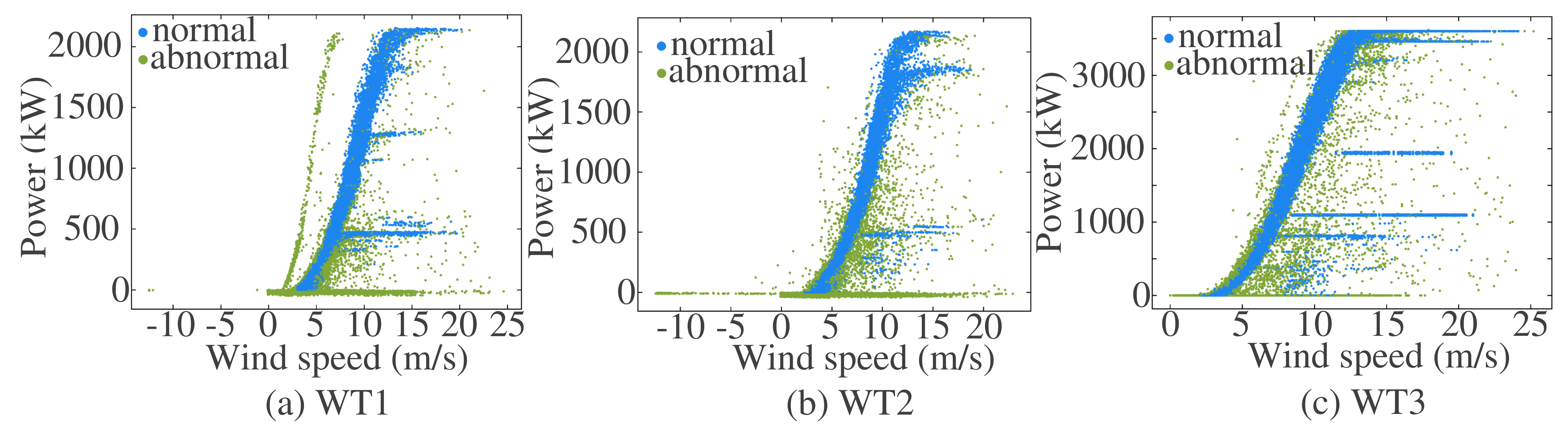}
    \caption{Outlier identification results by LOF.}
    \label{fig:enter-label9}
\end{figure}

\begin{figure}[!t]
    \centering
    \includegraphics[width=1\linewidth]{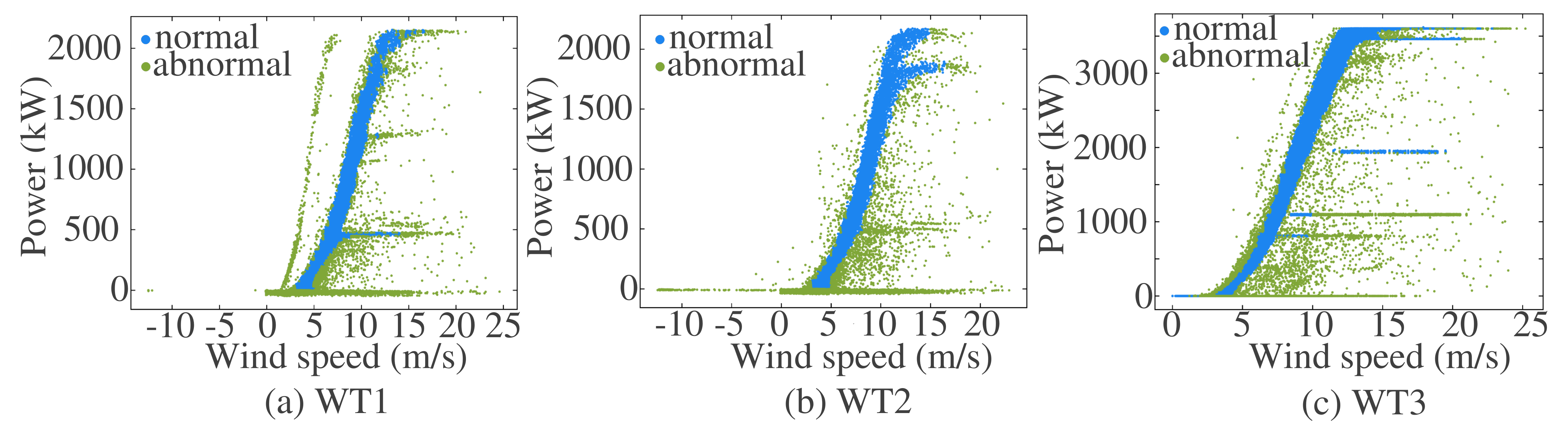}
    \caption{Outlier identification results by Quartile-DBSCAN.}
    \label{fig:enter-label10}
\end{figure}

\begin{figure}[!t]
    \centering
    \includegraphics[width=1\linewidth]{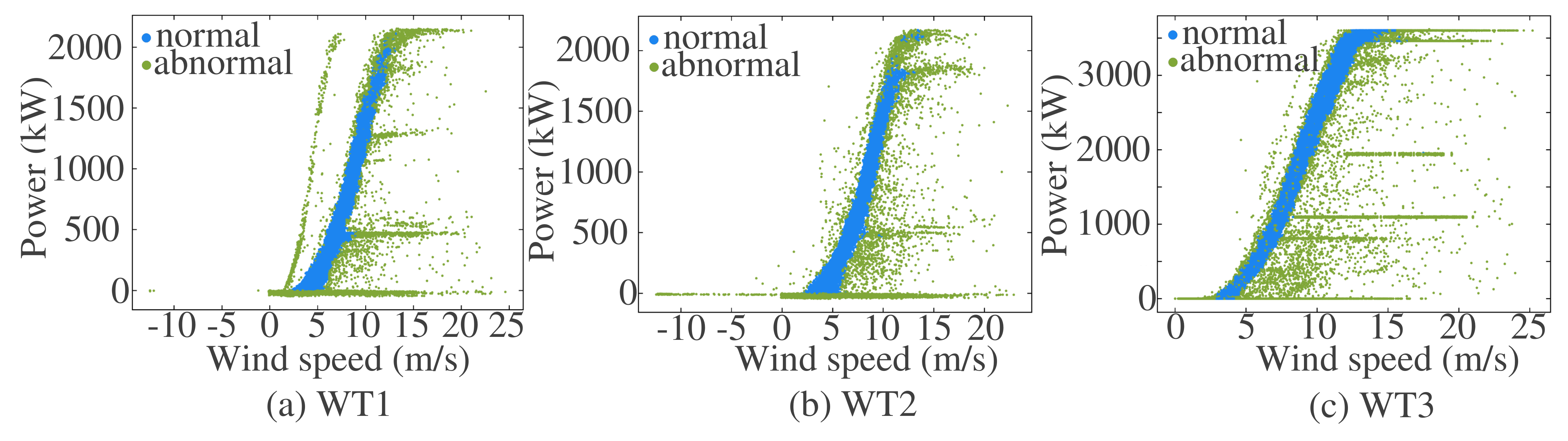}
    \caption{Outlier identification results by MDUE.}
    \label{fig:enter-label11}
\end{figure}

\begin{figure}[!t]
    \centering
    \includegraphics[width=1\linewidth]{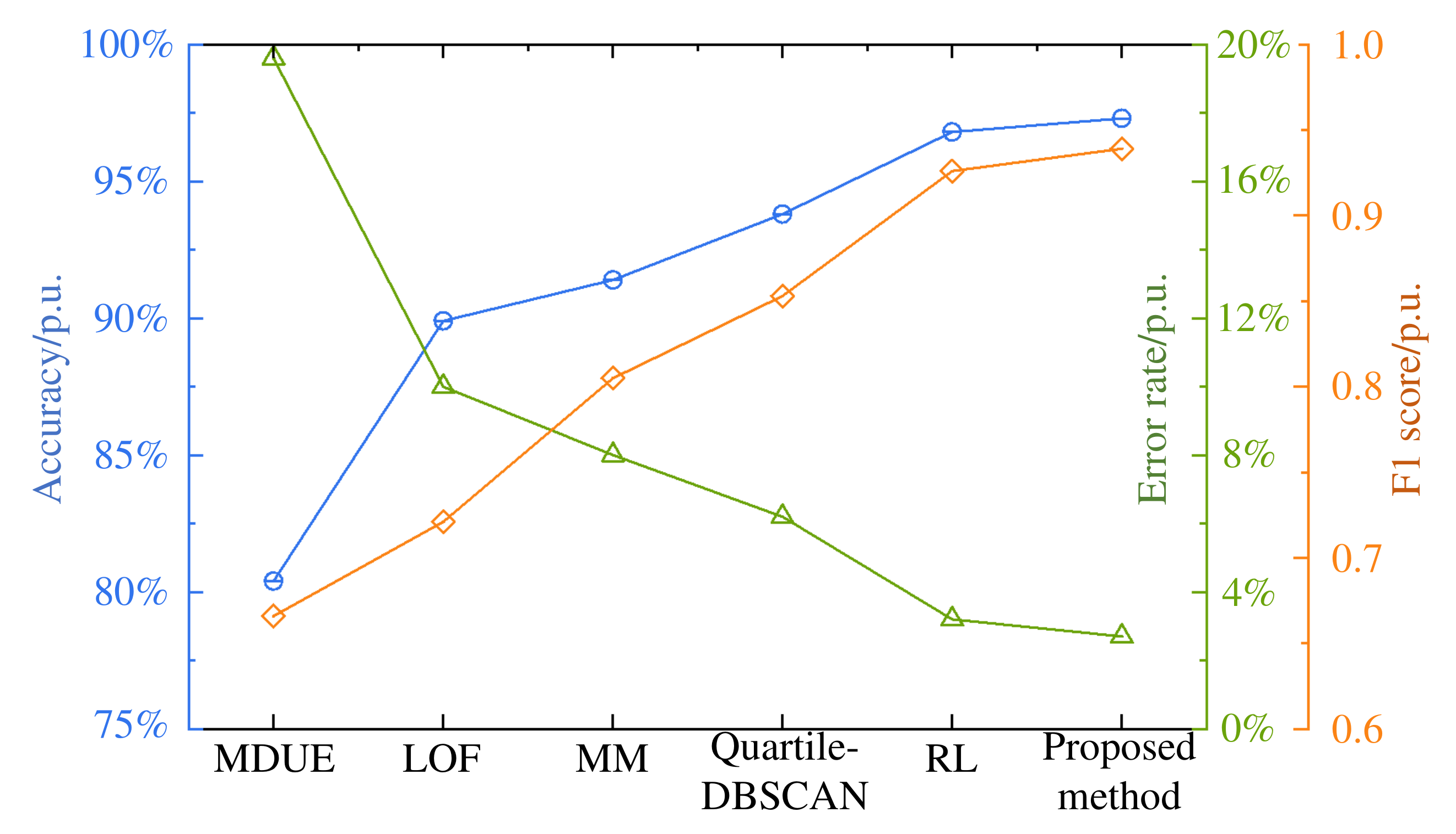}
    \caption{Identification metrics for various methods in WT3.}
    \label{fig:enter-label111}
\end{figure}

\subsubsection{Parameter Sensitivity Analysis}

For the three-stage outlier identification approach proposed in this paper, the critical parameters are the adjustment factor \textit{k} involved in the IQR method of the second stage and the diameter of the structuring element \textit{d} involved in the MMOs of the third stage. To systematically investigate the impact of these two parameters on the performance of the proposed methodology, parameter sensitivity tests are conducted here.

First, parameter sensitivity analysis for \textit{k} is implemented by setting its value to the range of 0.5$\sim$3.0 to check the outlier identification results (with \textit{d} fixed at 6). Taking the labeled dataset WT3 for instance, the parameter sensitivity test results are illustrated in Fig. \ref{k-sensitivity}. As can be seen, the anomalous data identification results of the proposed method are generally robust to the variation of \textit{k}, with satisfactory performance achieved when \textit{k} is set to 1.5 by default. 
Note that, as \textit{k} increases, the wind power fitting error $E_{\mathrm{rmse}}$ experiences a trivial variation, changing only from 0.029 to 0.031. Furthermore, for the accuracy and F1 score, they remain at relatively high levels of more than 96\% and 0.945 when $1 \leq k \leq 2.5$. Non-negligible performance degradation is observed when \textit{k} is excessively small or large. Therefore, to achieve desirable outlier detection performance, \textit{k} is recommended to be set to typical values of 1.0$\sim$2.5.

\begin{figure}[!t]
    \centering
    \includegraphics[width=1\linewidth]{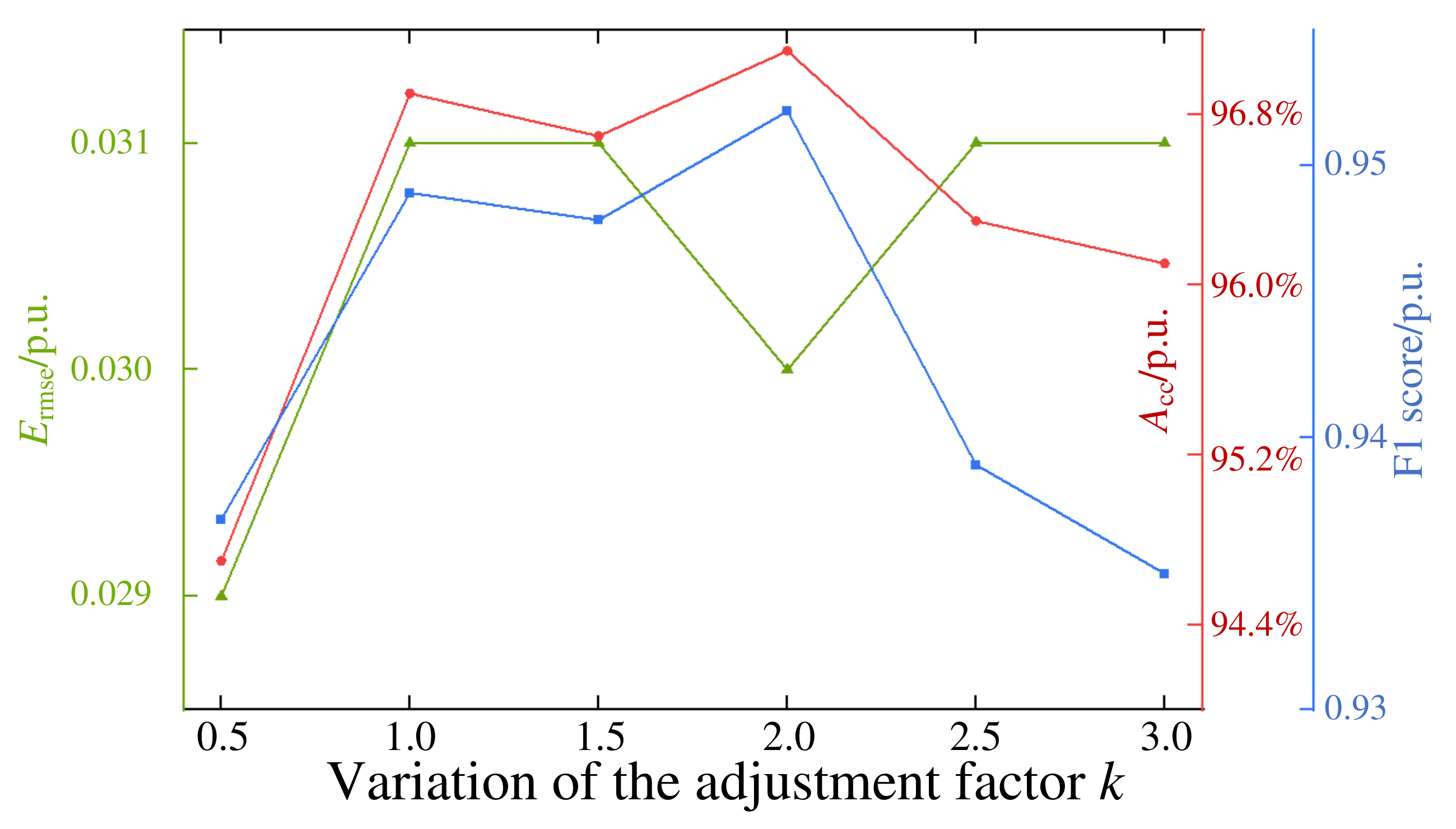}
    \caption{Outlier identification results for WT3 with the variation of the adjustment factor \textit{k}.}
    \label{k-sensitivity}
\end{figure}

With the same dataset (WT3), additional parameter sensitivity analysis is conducted for \textit{d} by varying it from 3 to 8 and observing the corresponding identification results (with \textit{k} set to 1.5), as shown in Fig. \ref{d-sensitivity}. From this figure, it is evident that when \textit{d} changes, the wind power curve fitting error $E_{\mathrm{rmse}}$ varies slightly, with the largest value being no more than 0.035. For the accuracy $A_{\mathrm{cc}}$ and the F1 score, they undergo gradual degradation with the increase of \textit{d}, especially when $d \geq 4$. In particular, both metrics decrease drastically when $d > 7$, with $A_{\mathrm{cc}}$ and the F1 score being no more than 94.5\% and 0.925. Overall, the outlier identification performance suffers from non-trivial degradation when \textit{d} is too large ($d > 7$) or too small ($d < 4$). Hence, based on the test results, it is advised to set \textit{d} to a moderate value within the interval of [4, 7].

\begin{figure}[!t]
    \centering
    \includegraphics[width=1\linewidth]{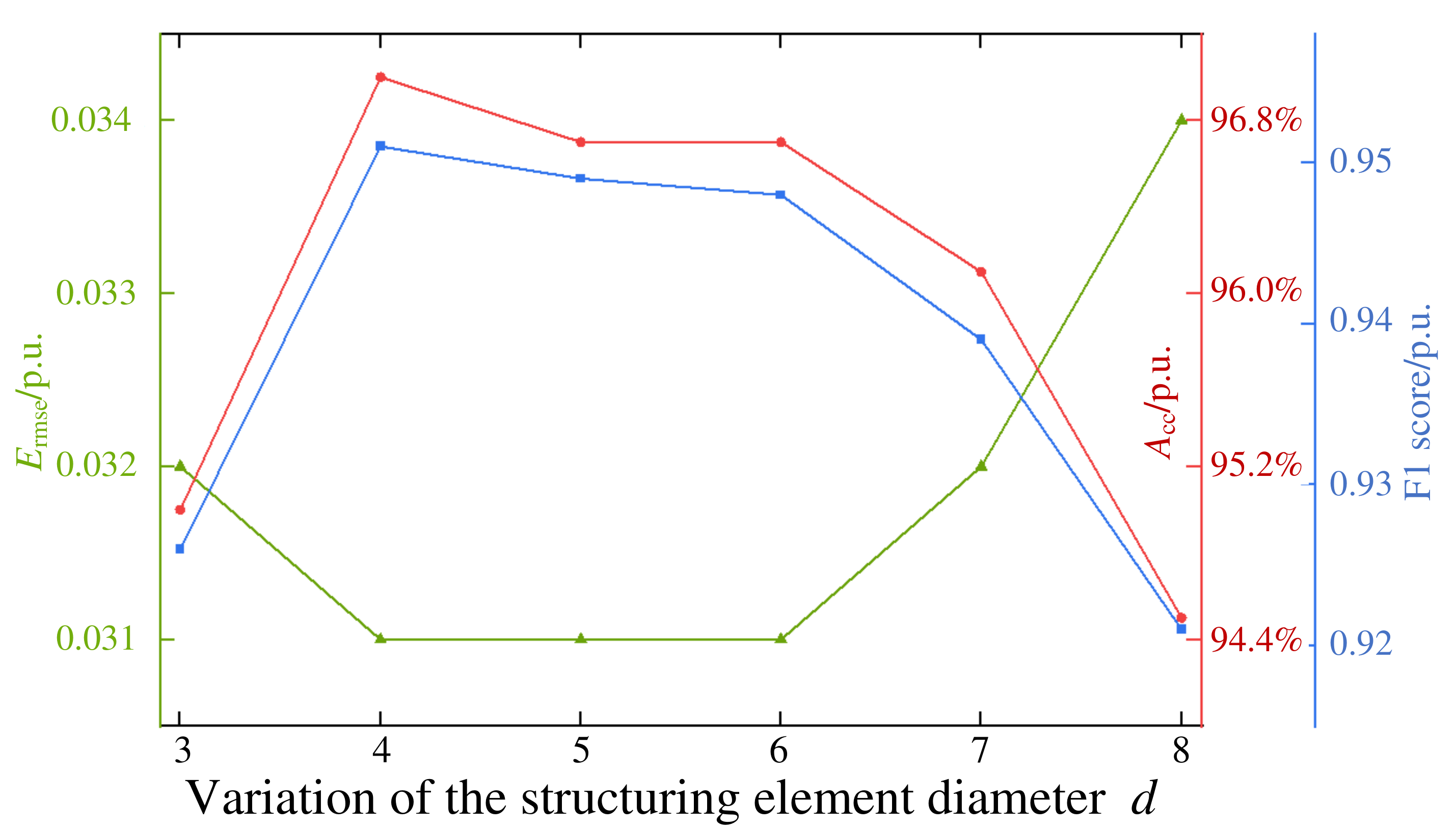}
    \caption{Outlier identification results for WT3 with the variation of the structuring element diameter \textit{d}.}
    \label{d-sensitivity}
\end{figure}

\subsubsection{Computational Efficiency}
To investigate the computational efficiency of different outlier identification methods, the time required to complete the outlier identification process for each method is examined. All the computations are implemented in Python and executed on a laptop equipped with a 3.10-GHz$*$16 Intel Core i5-12500H CPU and an NVIDIA GeForce RTX-3050ti GPU.

\begin{table}[!t]
    \belowrulesep=0pt
    \aboverulesep=0pt
    \renewcommand{\arraystretch}{1.3}
    \caption{Computation Times of Different Methods (Times in [\textit{s}]) }
    \label{tab4}
    \centering
    \begin{tabular}{c c c c c c c}
    \toprule
    \multicolumn{1}{c}{Dataset} & \makecell{Proposed \\ method} & \multicolumn{1}{c}{LOF} & \makecell{Quartile- \\ DBSAN} & \multicolumn{1}{c}{MDUE} & \multicolumn{1}{c}{RL} & \multicolumn{1}{c}{MM} \\
 \hline
    WT1                      & 5.57 & 2.73 & 3.68 & 28.7 & 2.47 & 2.79 \\
    WT2                      & 5.11 & 2.46 & 3.35 & 27.3 & 2.36 & 2.79 \\
    WT3                      & 5.90 & 3.87 & 4.52 & 39.2 & 2.99 & 3.03 \\
    \bottomrule
    \end{tabular}
\end{table}

As summarized in Table \ref{tab4}, the efficiency of the proposed method remains relatively high, completing the task of identifying all outliers in no more than 6 seconds. Although the computational efficiency of methods like LOF and RL is a bit higher than that of the proposed method, as illustrated above, the proposed method has higher reliability for outlier detection. Considering the dual demands on reliability and efficiency in practical applications, the proposed method would be more helpful in practice, enabling highly reliable online outlier detection without imposing a heavy computational burden.

\subsubsection{Potential in Enhancing Wind Power Prediction}
The main objective of outlier identification in this work is to improve the wind power data quality and provide reliable data sources for follow-up data-driven applications. Taking short-term wind power prediction as an application example, how the proposed method can help improve the performance of wind power prediction is tested here via comparisons. Specifically, the representative methods involved in Table \ref{tab888} are also implemented to examine their contribution to data quality enhancement. To ensure fair comparisons, this study employs a unified model based on a two-layer convolutional neural network (CNN) to realize short-term wind power prediction. This model is developed using the dataset initially processed by the outlier identification method, followed by cubic spline interpolation to fix outliers.
Then, the data are partitioned into training and test sets. Taking wind speed data in the next three days (for simplicity, the actual wind speed data are deemed as predicted values with no error) and power data acquired from the past three days as inputs, the power data for the next three days are predicted by CNN. In this manner, all outlier identification methods are examined, and the wind power prediction errors on test sets are presented in Table \ref{tab6}.

\begin{table*}[!t]
    \belowrulesep=0pt
    \aboverulesep=0pt
    \renewcommand{\arraystretch}{1.3}
    \caption{Errors of Wind Power Prediction Models Based on Different Identification Methods (Errors in [p.u.])}
    \label{tab6}
    \centering
    \begin{tabularx}{\textwidth}{c*{2}{>{\centering\arraybackslash}X}
                                     *{2}{>{\centering\arraybackslash}X}
                                     *{2}{>{\centering\arraybackslash}X}
                                     *{2}{>{\centering\arraybackslash}X}
                                     *{2}{>{\centering\arraybackslash}X}
                                     *{2}{>{\centering\arraybackslash}X}
                                     *{2}{>{\centering\arraybackslash}X}}
        \toprule
        \multirow{2}{*}{Dataset} 
        & \multicolumn{2}{c}{Unprocessed dataset} 
        & \multicolumn{2}{c}{Proposed method} 
        & \multicolumn{2}{c}{LOF}
        & \multicolumn{2}{c}{Quartile-DBSCAN}
        & \multicolumn{2}{c}{MDUE}
        & \multicolumn{2}{c}{RL}
        & \multicolumn{2}{c}{MM} \\
        & $e_{\mathrm{mae}}$ & $\mathrm{e}_{\mathrm{rmse}}$
        & $e_{\mathrm{mae}}$ & $\mathrm{e}_{\mathrm{rmse}}$
        & $e_{\mathrm{mae}}$ & $\mathrm{e}_{\mathrm{rmse}}$
        & $e_{\mathrm{mae}}$ & $\mathrm{e}_{\mathrm{rmse}}$
        & $e_{\mathrm{mae}}$ & $\mathrm{e}_{\mathrm{rmse}}$
        & $e_{\mathrm{mae}}$ & $\mathrm{e}_{\mathrm{rmse}}$
        & $e_{\mathrm{mae}}$ & $\mathrm{e}_{\mathrm{rmse}}$ \\
        \midrule
        WT1 & 0.0041 & 0.0064
             & 0.0032 & 0.0044
             & 0.0040 & 0.0061
             & 0.0038 & 0.0056
             & 0.0037 & 0.0052
             & 0.0035 & 0.0050
             & 0.0036 & 0.0052 \\
        WT2 & 0.0049 & 0.0069
             & 0.0038 & 0.0051
             & 0.0044 & 0.0061
             & 0.0042 & 0.0058
             & 0.0045 & 0.0062
             & 0.0041 & 0.0056
             & 0.0043 & 0.0060 \\
        WT3 & 0.0042 & 0.0057
             & 0.0034 & 0.0047
             & 0.0041 & 0.0055
             & 0.0039 & 0.0052
             & 0.0040 & 0.0053
             & 0.0038 & 0.0052
             & 0.0039 & 0.0051 \\
        \bottomrule
    \end{tabularx}
\end{table*}

As can be seen, the data quality is largely improved after the outlier identification, leading to a significant improvement in the accuracy of wind power predictions. The datasets processed by the proposed method achieve the highest prediction accuracy. Specifically, compared to models built upon unprocessed datasets, the accuracy of the proposed method exhibits an improvement of more than 20\%, whereas the remaining methods achieve enhancements by about 9\% to 15\%. This finding reveals the superiority of the proposed method in outlier identification, thereby significantly improving the data quality for wind power prediction. In this regard, the proposed method is capable of helping enhance the adaptability of subsequent data-driven applications under anomalous measurement conditions.

\subsubsection{Applicability to Other Datasets}
To examine the proposed method's applicability in the presence of other different datasets, it is tested on another three distinct wind power datasets: WT4, acquired from a wind farm in Ireland \cite{github}; WT5, acquired from a wind farm in China; WT6, obtained from numerical simulations in MATLAB/Simulink \cite{matlab}. Note that in WT4 and WT5, the outliers correspond to actual anomalies occurring in practical SCADA data, while in WT6, the outliers are produced by manually synthesizing data points significantly deviating from the simulation results.

In the raw SCADA data of WT4 and WT5, their historical operating logs are available to mark the abnormal operating status of practical wind turbines. Considering that such text information can be utilized to authentically annotate some outliers in the SCADA data due to abnormal wind turbine operations, the performance of the proposed method is examined by checking whether it can accurately detect such outliers. To clearly illustrate the performance, the outlier annotation results based on the operating logs and the outlier identification results of the proposed method are comparatively shown in Fig. \ref{wt4} and Fig. \ref{wt5}. As can be seen in Fig. \ref{wt4}(a) and Fig. \ref{wt5}(a), a significant portion of the anomalous data fails to be correctly identified by the logs, whereas the proposed method accurately identifies it, as marked by the black circles in the figures. For anomalous data identified through the logs, the proposed method can also detect them accurately. The $A_{\mathrm{cc}}$, $E_{\mathrm{rr}}$, and $F_{\mathrm{1}}$ are presented in Table \ref{tablog}.
As can be observed, the proposed method recognizes the majority of anomalous data identified through the logs, with only a few falsely dismissed. The F1 scores exceed 0.96, and $A_{\mathrm{cc}}$ across these datasets remains above 93.5\%, highlighting the method's excellent performance on outlier detection. Nevertheless, as mentioned above, the proposed method manages to identify a significant number of obvious anomalies not annotated by the operating logs. In this regard, it is not suggested to simply take outlier annotation results based upon operating logs as the only reference for performance validation.

\begin{figure}[!t]
    \centering
    \includegraphics[width=1.0\linewidth]{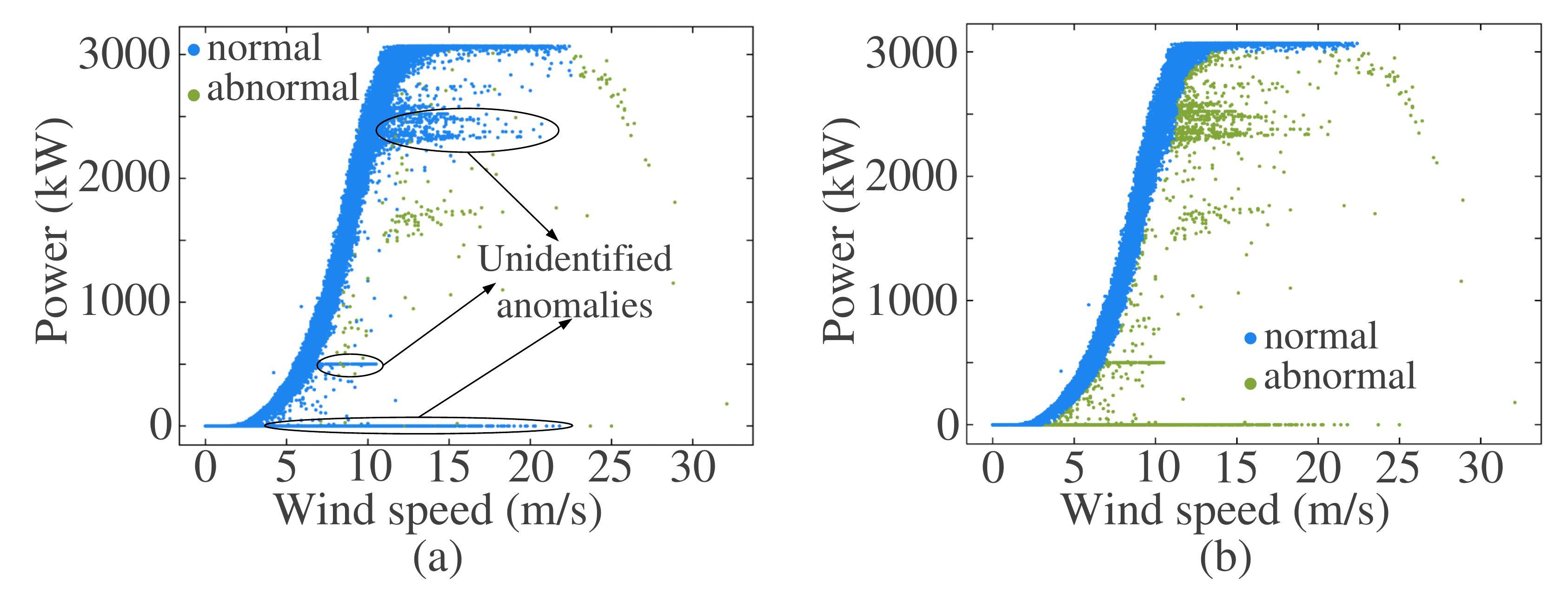}
    \caption{Outlier identification results for WT4. (a) Outlier annotation by checking operating logs. (b) Outlier identification based on the proposed method.}
    \label{wt4}
\end{figure}

\begin{figure}[!t]
    \centering
    \includegraphics[width=1.0\linewidth]{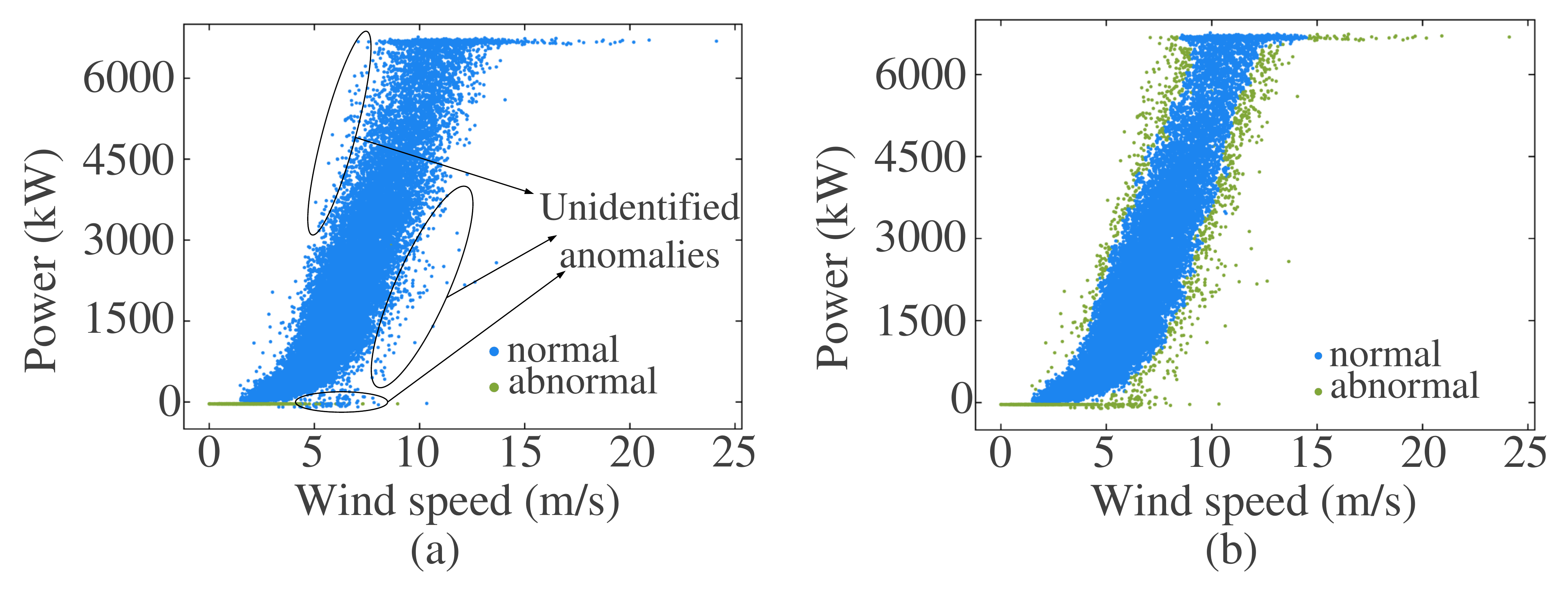}
    \caption{Outlier identification results for WT5. (a) Outlier annotation by checking operating logs. (b) Outlier identification based on the proposed method.}
    \label{wt5}
\end{figure}

\begin{table}[!t]
    \belowrulesep=0pt
    \aboverulesep=0pt
    \renewcommand{\arraystretch}{1.3}
    \caption{Effectiveness of The Proposed Method for Recognizing Anomalies Identified by Logs}
    \label{tablog}
    \centering
    \begin{tabular*}{\linewidth}{@{\extracolsep{\fill}}c c c c}
    \toprule
    Dataset & $A_{\mathrm{cc}}$/p.u. & $E_{\mathrm{rr}}$/p.u. & $F_{\mathrm{1}}$/p.u. \\ \hline
    WT4     & 93.7\% & 6.30\% & 0.968 \\
    WT5     & 100\%  & 0\%    & 1.000 \\
    WT5     & 100\%  & 0\%    & 1.000 \\
    \bottomrule
    \end{tabular*}
\end{table}

Further, to quantitatively show the capability of the proposed method in effectively detecting outliers dismissed by operating logs, the above-mentioned two types of metrics, i.e., the wind power curve fitting error ($E_{\mathrm{rmse}}$) and the wind power prediction error after outlier detection and correction ($e_{\mathrm{rmse}}$ and $e_{\mathrm{mae}}$) are used for comparative performance illustration, as presented in Table \ref{wt45}. As can be seen, the proposed method significantly reduces both the wind power curve fitting errors and the wind power prediction errors after identifying anomalous data. In contrast, since only a small proportion of anomalous data is annotated through the operating logs, the quality of the datasets simply cleansed by filtering out outliers marked via operating logs is much lower than that of the datasets processed by the proposed method, with much larger wind power curve fitting errors and wind power prediction errors. This again implies that wind turbine operating logs cannot be taken as a single reference to check if all outliers are correctly identified. In fact, this inadequacy indicates the necessity of systematic outlier identification schemes like the method proposed in this paper to filter out potential outliers that cannot be annotated by operating logs.

\begin{table*}[!t]
    \belowrulesep=0pt
    \aboverulesep=0pt
    \renewcommand{\arraystretch}{1.3}
    \caption{Outlier Identification Results for WT4 and WT5}
    \label{wt45}
    \centering
    \begin{tabularx}{\textwidth}{
        c*{3}{>{\centering\arraybackslash}X}
         *{3}{>{\centering\arraybackslash}X}
         *{3}{>{\centering\arraybackslash}X}}
        \toprule
        \multirow{2}{*}{Dataset} 
          & \multicolumn{3}{c}{Unprocessed dataset} 
          & \multicolumn{3}{c}{Identification through logs}
          & \multicolumn{3}{c}{Proposed method} \\
          & $E_{\mathrm{rmse}}$(p.u.)  
          & $e_{\mathrm{rmse}}$(p.u.) 
          & $e_{\mathrm{mae}}$(p.u.) 
          & $E_{\mathrm{rmse}}$(p.u.) 
          & $e_{\mathrm{rmse}}$(p.u.)
          & $e_{\mathrm{mae}}$(p.u.)
          & $E_{\mathrm{rmse}}$(p.u.)
          & $e_{\mathrm{rmse}}$(p.u.)
          & $e_{\mathrm{mae}}$(p.u.) \\
        \midrule
        WT4 & 0.130 & 0.0049 & 0.035
            & 0.127 & 0.0048 & 0.033
            & 0.027 & 0.0037 & 0.018 \\
        WT5 & 0.103 & 0.0050 & 0.031
            & 0.096 & 0.0048 & 0.031
            & 0.090 & 0.0040 & 0.023 \\
        \bottomrule
    \end{tabularx}
\end{table*}

Moreover, to verify its applicability in the presence of other data sources not collected from field measurements, the proposed method is applied to the dataset obtained from simulation, i.e., WT6, with the test results summarized in Fig. \ref{wt6} and Table \ref{rwt6}. From Fig. \ref{wt6}, it can be seen that the distribution of anomalous data identified through the proposed method highly resembles the actual distribution of original anomalies, with only a small number of points being misidentified. Table \ref{rwt6} also quantitatively shows that the proposed method achieves an excellent anomaly detection performance on this dataset, with the accuracy and the F1 score remaining over 94\% and 0.92, respectively. 
Furthermore, based on wind power fitting errors and prediction errors before and after the identification and correction of anomalous data, it can be concluded that the data quality of WT6 has been significantly improved.
Based on the above analyses, it is clear that the proposed method not only effectively handles datasets from different regions but also shows superior performance in recognizing anomaly data generated through simulations. This further demonstrates the robustness and superiority of the proposed method.

\begin{figure}
    \centering
    \includegraphics[width=1\linewidth]{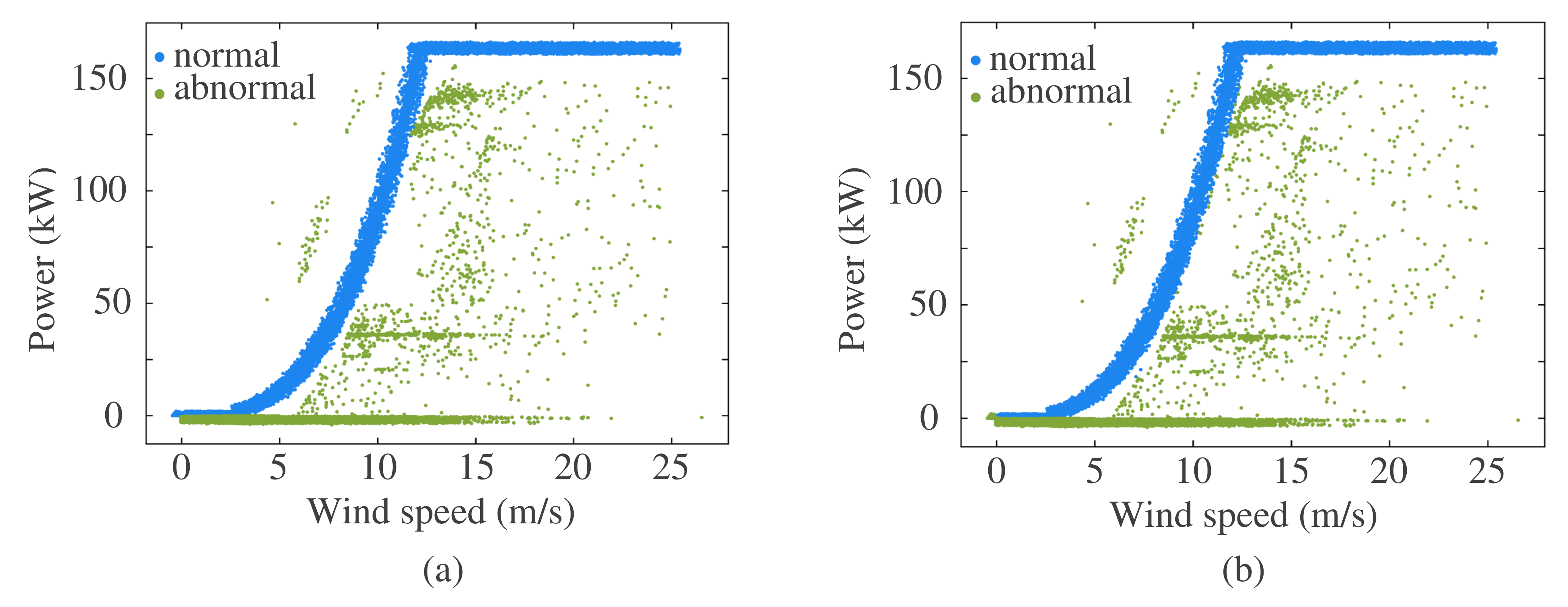}
    \caption{Outlier identification results for WT6. (a) True distribution of anomalous data. (b) Identification based on the proposed method.}
    \label{wt6}
\end{figure}

\begin{table}[!t]
    \renewcommand{\arraystretch}{1.3} 
    \caption{Outlier Identification Results for WT6 Based on The Proposed Method}
    \label{rwt6}
    \centering
    \begin{tabularx}{\columnwidth}{>{\centering\arraybackslash}X >{\centering\arraybackslash}X >{\centering\arraybackslash}X}
        \toprule
        Metrics & Unprocessed Dataset & Proposed Method \\
        \midrule
        $A_{\mathrm{cc}}$/p.u. & N/A & 94.7\% \\
        $E_{\mathrm{rr}}$/p.u. & N/A & 5.3\% \\
        $F_{\mathrm{1}}$/p.u. & N/A & 0.928 \\
        $E_{\mathrm{rmse}}$/p.u. & 0.156 & 0.013 \\
        $e_{\mathrm{rmse}}$/p.u. & 0.0044 & 0.0031 \\
        $e_{\mathrm{mae}}$/p.u. & 0.0031 & 0.0020 \\
        \bottomrule
    \end{tabularx}
\end{table}

\section{Conclusion}

This paper develops a three-stage composite outlier identification method to efficiently process wind power data. With the help of three complementary techniques, i.e., physical rule-based preprocessing, RL-enabled detection, and MM-based refinement, the proposed method is able to handle practical outliers with complicated distributions. Numerical test results on several real-world wind power datasets and a simulated dataset demonstrate that the proposed method is able to effectively identify both stacked and dispersive outliers in different wind power datasets, exhibiting significant outlier detection capability and desirable applicability in practical contexts. Compared to other representative methods, the proposed method is able to improve the outlier detection reliability without sacrificing the computational efficiency. On the basis of the proposed method, the performance of the subsequent application of wind power prediction can be significantly enhanced, with the prediction errors reduced by 20\%. Therefore, the proposed method shows a higher potential in enhancing wind power prediction, being more applicable in practical applications.

Note that, in addition to handling wind speed and power data, the proposed method can be extended to cope with outliers of other types of data acquired by wind turbine SCADA systems. Besides, given the high proportion of anomalous data in the dataset (more than 30\% in the above case studies), how to systematically fix the anomalous values is planned as a follow-up research direction of this work to ensure the data quality for practical SCADA data-enabled applications.


\appendices
\bibliographystyle{IEEEtran}
\bibliography{bibtex/bib/IEEEabrv,bibtex/bib/mylib}

%







\end{document}